# Reflexive Behaviour: How publication pressure affects research quality in Astronomy

**Julia Heuritsch**
Humboldt Universität zu Berlin, Research Group "Reflexive Metrics", Institut für Sozialwissenschaften, Unter den Linden 6, 10117 Berlin; julia.heuritsch@hu-berlin.de, julia.heuritsch@gmail.com

## Abstract

Reflexive metrics is a branch of science studies which explores how the demand for accountability and performance measurement in science has shaped the research culture in recent decades. Hypercompetition and publication pressure are part of this neoliberal culture. How do scientists respond to these pressures? Studies on research integrity and organizational culture suggest that people who feel treated unfairly by their institution are more likely to engage in deviant behaviour, such as scientific misconduct. By building up on reflexive metrics, combined with studies on the influence of organisational culture on research integrity, this study reflects on the research behaviour of astronomers: 1) To what extent is research (mis-)behaviour reflexive, i.e. dependent on perceptions of publication pressure and distributive & organisational justice? 2) What impact does scientific misconduct have on research quality? In order to perform this reflection, we conducted a comprehensive survey of academic and non-academic astronomers worldwide and received 3,509 responses. We found that publication pressure explains 19% of the variance in occurrence of misconduct and between 7 and 13% of the variance of the perception of distributive & organisational justice as well as overcommitment to work. Our results on the perceived impact of scientific misconduct on research quality show that the epistemic harm of questionable research practices should not be underestimated. This suggests there is a need for a policy change. In particular, lesser attention to metrics (such as publication rate) in the allocation of grants, telescope time and institutional rewards would foster better scientific conduct and hence research quality.

## 1. Introduction

The growing body of research on the effect of evaluation procedures on scientific behaviour (e.g. Hesselmann, 2014; Stephan, 2012; Laudel & Gläser, 2014; Fochler & De Rijcke, 2017) points towards performance indicators (such as publication and citation rates) not only describing, but also prescribing behaviour (Desrosières, 1998; Porter, 1995). In other words, they have *constitutive effects* on the knowledge production process (Dahler-Larsen, 2014). This suggests that metrics intended to measure concepts like research quality, end up *defining* what research quality means, and thereby *shaping* what researchers strive for. As Dahler-Larsen states: "A claim to measure quality cannot be understood as referring to an already-existing reality, but as an attempt to define reality in a particular way" (Dahler-Larsen, 2019; p.11). Metrics are therefore not merely proxies for quality, but represent a definition of what quality is considered as.

Capturing a complex concept such as research quality quantitatively, strips it of its complexity. This makes it easier to understand, turning it into something objective and comparable (Desrosières, 1998; Dahler-Larsen, 2019), but also leads to a "validity problem" (Dahler-Larsen, 2014; p.971). The indicator's inability to account for the phenomenon's full complexity may therefore lead to an "evaluation gap" (Wouters, 2017). This may lead to unintended consequences when putting indicators into place, such as *scientific misconduct* resulting from *coping* with the divergence between what quantitative proxies measure and what researchers value themselves (Heuritsch, 2021).

Which metrics are considered important depends on the *culture of science*, which has evolved through neoliberal reforms and the rise of New Public Management in the last 30 years (Lorenz, 2012). Researchers have become increasingly dependent on external resources, such as funding and rewards. *Competition* for these resources and for positions has intensified (Anderson et al., 2007). *Publish-or-Perish* is an integral part of this culture, since a scientist's reputation (arguably the most important currency in academia), their funding opportunities and their career development hinge on their metrics, such as their publication rate (Tijdink et al., 2014a; Moosa, 2018). As a result, researchers have an interest in scoring well on performance indicators. Due to this "goal displacement", where doing well on quantitative metrics becomes an aim in itself (Fochler & De Rijcke, 2017; p.27) researchers may adopt various gaming strategies to attain the goal. (refer to e.g. Laudel & Gläser, 2014; Rushforth & De Rijcke, 2015). This suggests that an indicator ceases to be a good measurement when it becomes a target (Goodheart's Law).

Adopting gaming strategies to hit a target set by performance indicators is what Fochler & De Rijcke (2017) call playing the "indicator game". While some forms of gaming may seem innocent at first (e.g. going for an easy publication), they can result in behaviour which scientists themselves perceive as a threat to *research integrity*. This may range from questionable research practices (QRPs), such as insufficient supervision of (graduate) students or salami slicing to publish more papers on one's research, to outright scientific misconduct such as fabrication, falsification or plagiarism (FFP; OSTP, 2000). Martinson et al. (2005) have shown that the latter, more extreme kinds of misconduct, are less frequent than the "'minor offences', the many cases of 'sloppy science'" (Haven & van Woudenberg, in print) and "carelessness" (Martinson et al., 2006; p.2), represented by the QRPs. Since they are more numerous and more difficult to spot, Martinson et al. (2005; p.737) suggested that such "mundane 'regular'" forms of misbehaviour pose larger threats to research integrity than outright fraud. If playing the indicator game is at least partly causing research misconduct, it follows that using indictors in research evaluation has an impact on *research integrity*.

Studies on the constitutive effects and unintended consequences of indicator use on research (behaviour) are designated under the umbrella term *"Reflexive Metrics"* (Heuritsch, 2019; p.146). The *relationship between research integrity* and *research culture & climate* has also been studied in literature (e.g. Crain et al., 2013; Martinson et al., 2013; Wells et al., 2014; Martinson et al., 2016). Academic culture may comprise of networks of peers, departments, institutions, funding agencies, grant reviewers, journal editors and the peer-review system (Martinson et al., 2006). These authors carried out the first systematic, quantitative analysis of the relationship between organisational culture, perceptions of justice and scientist's behaviours. They found evidence for greater perception of injustice leading to misbehaviour, especially among researchers whose career is at stake (e.g early-career researchers). Other studies related individual perceptions of research climate, such as advisor-advisee relations or expectations, to misconduct (e.g. Crain et al., 2013). Anderson et al. (2007) & Martinson et al. (2009) found evidence that the greater competition resulting from the neoliberalist culture in science resulted in gaming strategies to the detriment of research integrity.

Research integrity has been linked to *research quality* (e.g. Crain et al., 2013; Martinson et al., 2010), and the former is easier to measure than the latter (Wells et al., 2014). As both have a direct correlation with scientific misconduct, and a climate of research integrity fosters science quality, both terms are often equated in the literature on the effects of cultural aspects on research integrity. Crain et al. (2013; p.837) point out that improving research quality is in fact the "holy grail" of organizational initiatives targeted at an ethical organisational climate. Given that metrics have constitutive effects, simply "fixing" aspects of quality by an indicator (Dahler-Larsen, 2019; p.143) will likely not lead to this holy grail. Instead, quality indicators "are neoliberal instruments which colonize practices and undermine professional values" (Dahler-Larsen, 2019; p.14). Since directly measuring research quality therefore is unfeasible, studying cultural aspects about a research environment seems to be the way forward to foster scientific quality.

The effects of *publication pressure* on research quality have found particular attention in recent studies, since publication rate is one of the key metrics in academia (cf. Moosa, 2018; Stephan, 2012). Haven et al. (2019a) summarise some key studies' findings: while some extent of publication pressure may be a driver for productivity, too much of it may not only have negative effects on research integrity and quality, but also on individual researchers. Examples of these negative effects include secrecy (e.g. a lower willingness to share data; Zuiderwijk, 2019), less academic creativity, less reliable science, neglect of negative findings (Heuritsch, 2019 & 2021) and a greater likelihood to engage in misbehaviour (QRPs & FFPs; Bedeian et al., 2010; Tijdink et al., 2014b; Bouter, 2015). The perceived competition resulting from the *publish-or-perish* imperative may also lead to emotional exhaustion on the individual level and feelings of unworthiness (Tijdink et al., 2013; Tijdink et al., 2016; Heuritsch, 2019). However, previous studies on the effects of publication pressure on research quality have two shortcomings: 1) Quantitative studies have thus far mainly included scientists from specific disciplines, such as biomedicine, management and population studies and of specific academic ranks (Miller et al., 2011; Van Dalen & Henkens, 2012; Tijdink et al., 2014b). 2) While previous literature acknowledges the link between research integrity and research quality, we are currently lacking quantitative studies that explore the impact of misbehaviour on research quality as opposed to the implied impact on research quality through the compromised research integrity. 3) Haven et al. (under review) is the only quantitative study of our knowledge that studies both, the impact of publication pressure and the impact of research climate on research integrity.

To address these shortcomings, this paper aims to study quantitatively the impact of research culture on research quality in a natural science field – Astronomy. The aspects of research culture under scrutiny are: perceived publication pressure, perceptions of distributive justice and procedural justice in peer review, grant application and telescope time application processes. These cultural aspects have been found relevant in Astronomy in a qualitative study (Heuritsch, 2021). The author studies the "organisational hinterland" (Dahler-Larsen, 2019) of Astronomy, to understand how quality inscriptions are produced, how they diverge from the astronomers' definition of quality and how this discrepancy affects research behaviour. In a nutshell, Heuritsch (2021) finds evidence for the structural conditions in the field, especially the over-emphasis on performance measured by publication rate, reception of external grants and telescope time, leading to gaming strategies to score well on those indicators mostly in the form of QRPs. These are found to be a response to the dissonance between cultural values (producing qualitative research that genuinely pushes knowledge forward) and the institutional objectives imposed to have career in academia (scoring well on indicators). In other words, there is a discrepancy between what indicators measure and the astronomers' definition of scientific quality – the so-called evaluation gap. Gaming strategies then give the appearance of compliance with cultural values, while using institutionalised means to achieve a good bibliometric record in innovative ways, such as salami slicing, cutting corners or going for easy publications (Haven et al., 2019b). The author finds evidence for a decrease in overall research quality as a consequence of prioritising quantity.

Based on Heuritsch (2019 & 2021) we can use astronomers' own definition of research quality, as well as previous studies on the relationship between academic culture and research behaviour, to analyse the effect of perceived publication pressure and organisational justice on research behaviour **and** -quality in astronomy. Understanding which cultural aspects foster and which inhibit research quality in this field will bring us a step closer towards the holy grail – the knowledge of how to support the scientific enterprise.

This is not only the first study to analyse the effects of cultural aspects (such as publication pressure) on research quality in Astronomy, but also the first to integrate Reflexive Metrics and studies on the relationship between research culture and -integrity. Moreover, it is the first to employ structural equation modelling to fully account for the structural relationships between the phenomena of interest.

This Paper will be structured as follows. First, we will give a theoretical background on explanations of misconduct, which contains a review on research integrity studies taking organisational culture into account. Second, the method section describes the sample selection, the survey instruments, research question & hypotheses and technical aspects with regards to how we performed the statistical analyses. Third, the result section contains descriptive statistics, the results from our EFAs, CFAs and SEM with scientific misconduct as the dependent variable, as well as the perceived impact of scientific misconduct on research quality. The result section is followed by a discussion, strength & limitations and a conclusion section that also gives an outlook for future studies.

## 2. Theoretical Background: Explanations for Misconduct

To understand metrics' role in research misconduct (hereafter referred to as *misconduct* or *misbehaviour*), we must first reflect on the causes of misbehaviour. Haven & van Woudenberg (in print), based on Sovacool (2008), suggest that there are three narratives which can explain misconduct:

(i) Failures on the individual actor's level ("Impure individuals")
(ii) Failures on an institutional level (of a particular university/ institute)
(iii) Failures on the structural system of science level

Haven & van Woudenberg (in print) point out that these narratives are not mutually exclusive, and tests six theories taken from previous literature on misconduct to assess their value in explaining one or more of these narratives. Five of those six theories shall be considered for this study:
(1) Bad Apple-,
(2) General Strain-,
(3) Organizational Culture-,
(4) New Public Management and
(5) Rational Choice Theory.

While a comprehensive review of these theories is outside the scope of this paper, more background information on this discourse can be found in Martinson et al. (2006) and Haven &

van Woudenberg (in print). To set the theoretical background for this story, we shall give a brief overview and describe how (1) to (4) can be subsumed under (5) – Rational Choice Theory.

**Bad apple theories** provide perhaps the earliest explanations for misconduct (Hackett, 1994) and account for the first narrative (i), since misbehaviour is thought to be solely caused by an individual and their distorted psychology. However, these theories are regarded as too simplistic in a sociological context, as they do not account for any institutional. (ii) or structural. (iii) contexts (Martinson et al., 2013).

**General Strain Theory** (GST; Agnew, 1992) is an "important strand of deviance theory, as it is the pressure to deviate from accepted norms as a response to perceived injustice" (Martinson et al., 2006; p.4). Based on Durkheim's concept of anomie, Merton (1938) suggests that deviant behaviour may be a *coping response* to structural strain, resulting from the inability to meet cultural ends with culturally legitimate means. The deviant behaviour (such as scientific misconduct) motivated by this kind of stress, is therefore nothing else but an "innovative" pathway to success. Agnew's GST enhances Merton's concept with the idea that deviant behaviour is not a necessary outcome of strain, but that coping strategies also depend on individual traits, such as self-esteem and intelligence, and contextual factors, such as a strong social support network and whether peers show legitimate or illegitimate coping behaviour. Since GST recognises individual-environment interactions and strain resulting from structural conditions, it accounts for all three narratives.

**Organisational Culture Theories** (OCTs), rooted in organisational psychology, recognise that the culture and structure of the organisation an individual works in affects their behaviour. In the context of academia, the organisational structure spans local. (e.g. departments or research institutes) and external settings (e.g. funding agencies, [inter-]national peer review systems, the overall academic employment market, etc; Martinson et al., 2006). Organisational culture encompasses all explicit and implicit norms and values within the organisation. A particular strand of OCT, Organisational Justice Theory (OJT), suggests that individuals who perceive they are being treated fairly by their organization, behave more fairly themselves (Martinson et al., 2006[1]; Martinson et al., 2010). The fairer people feel their organization's processes are, the more likely they are to trust their workplace, to comply with decisions made and to not engage in questionable behaviour (ibid.). In other words, people who perceive the distribution of resources and decision making processes as fair are more likely to respond with normative as opposed to deviant behaviour, such as scientific misconduct (Martinson et al., 2006). One may distinguish between two types of organisational justice (OJ; ibid.): procedural and distributive justice. The former refers to a perception of fairness in decision-making and the latter to fairness in resource distribution processes. In academia, these processes may stem from the local and external settings of the organisational structure, such as peer-review of manuscripts, tenure, promotion and peer-review committees for research grant proposals (Martinson et al., 2010). Since OCTs recognise the fact that characteristics of the environment in which researchers work promote or inhibit scientific integrity, they are of type institutional. (ii) and structural. (iii) narratives.

**New Public Management** (NPM) is a form of public administration based on neoliberal policies and is characterised by a combination of free market ideology and intense managerial control practices (Lorenz, 2012). The author (p.601) poses the formula: "free market = competition = best value for money = optimum efficiency […]". Arguably, this formula is the rationale for competition in science, since NPM practices have reached academia since the

[1] These authors also give more information on the development of OCTs and OJTs, which would go beyond the scope of this paper.

1980s (ibid.). The NPM paradigm also values efficiency as a key objective, as evident from the formula. The call for increased accountability in science (cf. Espeland & Vannebo, 2008) can be associated with this strive for efficiency. Accountability is in turn sought using quantitative performance indicators, such as publication rates and impact factors. Since resources are limited, there are less tenured positions in science than graduate students and given the extreme focus on efficiency and performance, NPM may result in "hypercompetition" in academia (Halffman & De Radder, 2015) at the cost of research integrity and the scientific enterprise (Anderson et al., 2007).

> *"Academic misconduct is considered to be the logical behavioral consequence of output-oriented management practices, based on performance incentives."* (Overman et al., 2016; p.1140).

Therefore, the NPM theory of scientific misbehaviour suggests that if there is an over-emphasis on performance and competition, researchers will tend towards "self-protective and self-promoting" behaviours such as mistrust in peers, an aversion towards sharing information & data , QRPs and FFPs (Anderson et al., 2007; p.459). This theory falls under the structural system of science narrative (iii).

**Rational Choice Theory** (RCT), is proposed by Heuritsch (2021), as a suitable framework to study the emergence and the impact of the evaluation gap in the academic field of Astronomy. Despite RCT's roots in economics, and contrary to often stated in literature (e.g. Haven et al., under review; Atkinson-Grosjean & Fairley, 2009), when applied thoroughly in sociology (according to Esser, 1999), RCT does not suggest that individuals act "rationally" in the classical sense (i.e. high investment of cognitive resources and disregarding emotions/ instincts) or the simplistic "Homo Economicus". Instead, RCT explains how it is often more rational not to invest cognitive resources and instead follow scripts, which are pre-defined instructions on how to act in situations based of cultural and/ or institutional norms (Esser, 1999). Thorough application of RCT follows the Coleman Boat (Coleman, 1990) by first analysing the *logic of an actor's situation*, which is comprised of one internal and three external constituents: (1) the internal component, made up of the actor's values, drivers, skills and personality; (2) the material opportunities at present; (3) explicit and implicit institutional norms and (4) the cultural reference frame, such as symbols and shared values. This *situation* is in a second step translated into *bridge hypotheses* which deliver the variables that are relevant for the individual's action that follows from the *situation*. An *action theory* (oftentimes the Expected Utility Theory; cf. Esser, 1999 and Heuritsch, 2021; Fig 3) is applied to explain how choices are made based on the derived variables of the *situation*. Third, by finding suitable *transformation rules* one can explain how individual actions aggregate to the sociological phenomenon in question. The result is a sociological explanation of how the interplay between structural conditions, institutional norms and the individual's personalities cause collective social phenomena. Applied to explaining scientific misbehaviour, RCT therefore pays tribute to all three narratives.

Building up on Heuritsch (2021), which followed the Coleman Boat in their analysis, we find that theories (1) to (4) may be subsumed under RCT. First, the internal constituent of the individual's *situation* accounts for the person's beliefs, values and drivers. If they do not correspond to the cultural values, such an individual may assume the role of a "bad apple" (1) in the respective system/ organisation. Second, if the cultural values cannot be lived up by institutionally legitimate means, the individual may perceive strain in the form of anomie (2; cf. Heuritsch, 2021). Third, the external constituents of the *situation* account for the organisational culture and environment (3), and therefore its subsequent relevance in the individual's action. For example, perceived injustices may contribute to strain (2; Martinson et al., 2010). Fourth, the prevailing NPM paradigm (5) influences the organisational culture (3), and therefore explains part of the culture's values and norms (Haven et al., under review). For example, NPM is likely to foster hypercompetition, which may impose strain (2), which in turn may lead to deviant behaviour, depending on individual's dispositions (1) that regulate the response to strain. This is how theories (1) to (4) deliver partial explanations for misbehaviour, which subsumed under RCT (5) can achieve a wider explanatory value.

## 3. Methods

### 3.1 Sample Selection and procedure

Astronomy is a highly international and globalised field (Roy & Mountain, 2006; Heuritsch, 2021). This includes a high proportion of international collaborative research (Chang & Huang, 2015), not least because of the sharing of observatories located in specific parts of the world (e.g. ALMA in Chile). Astronomers have a strong common culture through their publication systems (three main journals and conference proceedings), societies and professional associations (Roy & Mountain, 2006; Heuritsch, 2021). Heidler (2011), estimates there to be about 15,000–20,000 active astronomers worldwide. They may work in academic research or in other research facilities, such as space agencies or non-public institutes. Given the close similarity in culture of academic and non-academic astronomers, this survey targeted both groups.

The ideal aim was to run a census. However, there is no official, complete list of all astronomers worldwide. Therefore, we used a three-stage cluster sampling technique to build our sampling frame, encompassing as many astronomers as possible. In the first stage, we constructed a list of astronomy institutions worldwide, including universities, non-academic research organisations, observatories, societies and associations. In the second stage, based on this list, we reached out to 176 universities, 56 non-academic research facilities & observatories and 17 societies & associations. We did so by emailing a respective contact person (e.g. a secretary/ department head), which included an invitation to the survey, asking them to forward the email to all department members (including PhD students). We estimate that 1,200 academic astronomers were reached in this way. In the third stage, we contacted the heads of the 9 divisions[2] of the International Astronomical Union – the largest association of academic and non-academic astronomers – asking them to forward the invitation to their respective division members. Three of them followed our request and one posted the invitation in their newsletter. In the fourth and final stage, we used an automated script to send out email invitations to the five remaining divisions, whose their heads were non-responsive, using publicly available email adresses. That way all IAU members[3] (approximately 12000 astronomers) were reached through at least one channel. Note, that some astronomers may have received the invitation more than once, since astronomers may be part of more than one division and may also have been reached through multiple of the approaches described above. Given that the IAU only has around 7% of "junior members", we asked survey recipients to forward the invitation to early career researchers. We estimate that around 13000-15000 astronomers were reached in total, 3509 of which at least party completed the survey, amounting to a response rate of roughly 25%. 2011 astronomers completed the survey in full.

[2] https://www.iau.org/science/scientific_bodies/divisions

[3] https://www.iau.org/administration/membership/member/

### 3.2 Instruments

We used the online tool LimeSurvey to create and host the survey. As outlined above, the survey is embedded in the conceptual framework of rational choice theory (RCT), subsuming organisational culture theory (OCT) and its subtype organisational justice theory (OJT). Therefore, our online survey contains a number of instruments to measure our independent and dependent variables:

#### Dependent Variables

- Scientific (Mis-) Behaviour
  While there is no general definition of research misconduct, since it is highly contextual (Hesselmann, 2014), Martinson et al. (2005, 2006, 2009, 2010) and Bouter et al. (2016) designed questionnaires asking about the occurrence of, in total, 90 different misbehaviour and questionable research practices. On the basis of Heuritsch (2021) we chose for our study the 18 items that were most relevant for the context of astronomy (see *S1-Appendix: S1-Table1a-b*). In our survey, we ask about the perceived frequency of observed (mis-)behaviour.

- Research Quality
  We operationalised research quality in astronomy on basis of the findings of Heuritsch (2019 & 2021). The author found three quality criteria: (1) good research needs to push knowledge forward, which includes studying a diversity of topics and making incremental contributions; (2) The research needs to be based on clear, verifiable and sound methodology that is (3) reported in an understandable and transparent way. This includes the sharing of data and reduction code. In the line of Bouter et al. (2016) who surveyed the frequency of misbehaviour and its impact, for each (mis-) behaviour item we asked about the frequency (as mentioned above), the impact on the validity of the findings, the impact on the communication value of the resulting paper and for two items we additionally asked for the impact on the research diversity (see *S1-Table1a-b*). In analogy to Bouter et al. (2016; p.2), “the total harm [on quality] caused by a specific research misbehaviour depends on the frequency of its occurrence and the impact [on quality] when it occurs”.

#### Independent Variables

- Perceived Publication Pressure
  To measure perceived publication pressure, which has been linked with (perceived) misbehaviour (Haven et al., under review), we adapted the Publication Pressure Questionnaire (PPQ; as validated by Tijdink et al., 2014a) to the context of research in astronomy. The initial PPQ consists of 18 items and we added four more. The added questions dealt with the influence of perceived publication pressure on the publication of data, reduction algorithms and replicability – all three of which have been found important for research quality in Astronomy (Heuritsch 2019 & 2021). The 22 adapted PPQ items can be found in *S1-Appendix: S1-Table2*.

- Perceived Organisational Justice: Distributive & Procedural Justice
  We measure perceived distributive justice via the Effort-Reward Imbalance (ERI; Siegrist et al., 2014) instrument. We adapted the short version, which consists of 10 items and added one more item, resulting into three effort and eight reward items (see *S1-Appendix: S1-Table3*). With regards to perceived procedural justice, in this study we consider the following processes: a) Resource allocation, b) Peer review, c) Grant application and d) Telescope time application. We adapted the instruments Martinson et al. (2006) used to the context of astronomy and the specific processes (see *S1-Table4a-d*). In particular, we added questions about how much the success of that process depends on luck on the one hand and improper preferential treatment on the other hand. The addition of these two items followed from findings by Heuritsch (2021) and suggestions from initial tests with astronomers.

- Perceived Overcommitment
  Perceived overcommitment (OC), the inclination to overwork, may be positively related to misbehaviour (Martinson et al., 2006). We adapted the six overcommitment items contained in the ERI instrument (Siegrist et al., 2014) (see *S1-Appendix: S1-Table5*).

### Control Variables

In addition to our main independent variables that may predict misbehaviour, we assume that role-associated and individual aspects factor into these relationships. On basis of Heuritsch (2021), we expect that scientists who haven't yet established their reputation, such as early career researchers, will be more likely to perceive publication pressure and organisational injustice, as there are insufficient tenured positions. As opposed to Martinson et al. (2006), who suggests that role-associated aspects ("social identity") mediate the relationship between the independent and the dependent variables, we therefore propose that they predict the independent variables. Our control variables include: gender, academic position, whether one is primarily employed at an academic (as opposed to non-academic) institution, whether one is employed at an institution in the global North/ South[4], and number of published first or co-author papers in the last 5 years. The reference categories are as follows: gender: female/non-binary; academic position: full professor; primary employer: Non-Academic; institute location: Global South; number of papers published: 1–5.

## 3.3 Research Question & Hypotheses

In light of the theoretical background as outlined above, we work from the assumption that higher perceived procedural and distributive injustice research and higher perceived publication pressure, will increase scientists' chances to observe research misbehaviour. We expect early-career researchers, with a less secure position, to be more likely to perceive both injustice and publication pressure.

Our research question is: To what extent can role-associated factors, cultural aspects and publication pressure explain the variance in perceived research misbehaviour and what effect does misbehaviour have on the research quality in astronomy?

Building on the qualitative study of the evaluation gap and its potential consequences on research quality in astronomy by Heuritsch (2021) and previous studies on the relationship between research culture and integrity (e.g. Martinson et al., 2006), this study tests the following hypotheses by means of a quantitative survey:

[4] https://meta.wikimedia.org/wiki/List_of_countries_by_regional_classification

1. (H1): The greater the perceived *distributive injustice* in Astronomy, the greater the likelihood of a scientist observing misbehavior.
2. (H2): The greater the perceived *organizational injustice* in Astronomy, the greater the likelihood of a scientist observing misbehavior.
3. (H3): The greater the perceived *publication pressure* in Astronomy, the greater the likelihood of a scientist observing misbehavior.
4. (H4): Those for whom injustice and publication demands pose a more serious threat to their academic career (e.g., early-career & female researchers in a male-dominated field) will perceive the organizational culture to be more unjust, the publication pressure to be higher, and subsequently more occurring misbehavior.
5. (H5): Scientific misbehavior has a negative effect on the research quality in astronomy.
6. (H6): The greater the perceived *publication pressure* in Astronomy, the greater the likelihood of a scientist perceiving a greater distributive and organizational injustice.

Based on these hypotheses, we specified our structural equation model. Given that publications are the main output of research and one of the key indicators used to evaluate the performance of astronomers (Heuritsch, 2021), we hypothesise that the perception of distributive and organizational justice depends on the perceived publication pressure. Overcommitment to work may depend on the perceived publication pressure as well as perceived distributional and organisational justice. We test the influence of all our control variables (academic position, gender, number of published papers, being academic/ non-academic, employment in global north/ south) on all our independent variables and the dependent variable, scientific misconduct. *Figure 1* depicts the level of the latent constructs of this model, excluding the measured indicators, for simpler readability.

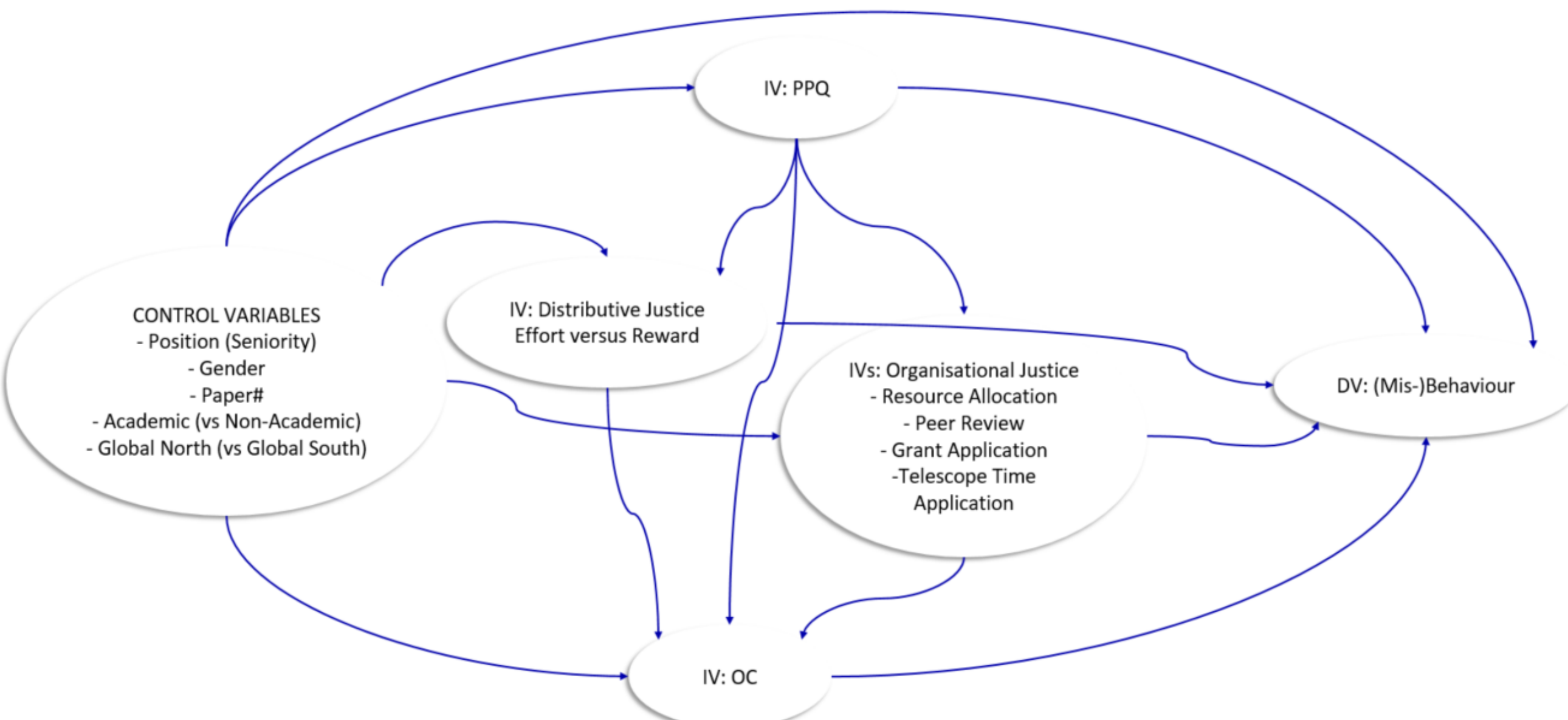

*Figure 1: Visualisation of our SEM on the level of latent constructs of our control, independent (IV) and dependent variables (DV). Arrows indicate that a construct (arrow head) is being regressed onto another construct (arrow start).*

3.4 Statistical Analyses

The analysis of this survey was performed in SPSS and R. Data preparation, including recoding and calculation of mean scores was performed in SPSS. We decided to exclude the 23 Bachelor and Master students from our sample, since we received too few responses from this category to conduct a proper analysis. All instruments measuring the independent variables (see *S1-Appendix*), are scored on a scale from 1 (strongly disagree) to 5 (strongly agree) and are treated as continuous variables in our structural equation model (SEM). The steps to arrive at our final model started with testing the independent variable constructs PPQ and ERI (including overcommitment) by performing an exploratory factor analysis (EFA) for ordinal data (CATPCA in SPSS) and we derived Cronbach Alphas as scale reliabilities. For the EFA we used Promax Kaiser-normalisation for rotating the factors. Next, all independent variable constructs were tested by means of confirmatory factor analysis (CFA) using Lavaan version 0.5-23 (Rosseel, 2012) in R version 3.3.1. For each construct, we used the residual correlation matrices to determine significant correlations of the indicators and included them into the respective models. After checking for construct validity, we further used Lavaan to perform structural equation modelling, which is the purpose it was designed for. Lavaan uses maximum likelihood estimation for regression analysis and listwise exclusion for missing data. The results section will present the results of our EFAs, CFAs and the complete SEM.

# 4. Results

## 4.1 Descriptive Statistics

In *Table 1* we first present the descriptive statistics of the control variables. Females make up around 26% of the sample (N=1827). For further analysis we combined the female and non-binary categories. Out of 2188 astronomers who shared their academic position in the survey, there are about 15% PhD candidates, 23% postdocs, 8.5% assistant professors, 25.96% full professors and 12.25% unranked astronomers. Out of the 2478 astronomers who declared whether their primary employment is at an academic university setting, 84.18% are employed in such setting and 15.82% are employed at other institutions which do research, such as national research institutes, observatories/telescopes or space agencies. Out of 1624 astronomers who answered in which country their primary employment is located, 84.79% astronomers work for an institution in the global North and 15.21% in the global South. 2610 astronomers answered how many papers they published as first or co-authors in the last 5 years. The 7.8% who have not published any yet are excluded from our regression analysis, since they did not receive the item battery regarding organisational justice in terms of peer review. The largest publication category is 1-5 papers published in the last 5 years with 31.72% of respondents, followed by the 11-20 publications category with 16.21% of respondents. 3.53% have been first or co-author for more than 100 papers in that time frame. Out of 2647 astronomers who answered the question of whether they applied for telescope time in the past 5 years, 58.9% replied yes and the rest replied no. The same amount of people filled in the question about whether they applied for a grant application in the past 5 years, and here 62.3% answered yes. Those who answered no for any of the two questions did not receive the item batteries regarding procedural justice with respect to telescope time or grant application processes, respectively.

| Baseline characteristic | n valid | Frequency | Percent | Valid Percent |
|---|---|---|---|---|
| Gender | 1827 | | | |
| Male | | 1333 | 38 | 73.0 |
| Female | | 482 | 13.7 | 26.4 |
| Non-Binary | | 12 | 0.3 | 0.7 |
| Academic Position | 2188 | | | |
| PhD candidate | | 332 | 9.5 | 15.2 |
| Postdoc/ Research Associate | | 504 | 14.4 | 23.0 |
| Assistant professor | | 186 | 5.3 | 8.5 |
| Associate professor | | 330 | 9.4 | 15.1 |
| Full professor | | 568 | 16.2 | 26.0 |
| Other | | 268 | 7.6 | 12.2 |
| Academic/ Non-Academic | 2478 | | | |
| Academic | | 2086 | 59.4 | 84.2 |
| Non-Academic | | 392 | 11.2 | 15.8 |
| Location of Employment | 1624 | | | |
| Global North | | 1377 | 39.2 | 84.8 |
| Global South | | 247 | 7 | 15.2 |
| Numbers of published papers (as first or co-author during the past 5 years) | 2610 | | | |
| 1st paper currently under review | | 54 | 1.5 | 2.1 |
| 0 | | 150 | 4.3 | 5.7 |
| 1-5 | | 828 | 23.6 | 31.7 |
| 6-10 | | 389 | 11.1 | 14.9 |
| 11-20 | | 423 | 12.1 | 16.2 |
| 21-30 | | 173 | 4.9 | 6.6 |
| 31-40 | | 158 | 4.5 | 6.1 |
| 41-50 | | 131 | 3.7 | 5.0 |
| 51-60 | | 61 | 1.7 | 2.3 |
| 61-70 | | 43 | 1.2 | 1.6 |
| 71-80 | | 40 | 1.1 | 1.5 |
| 81-90 | | 31 | 0.9 | 1.2 |
| 91-100 | | 37 | 1.1 | 1.4 |
| >100 | | 92 | 2.6 | 3.5 |

*Table 1: Descriptive statistics of the control variables of N=3509 survey respondents*

For each independent and dependent variable construct, we calculated the mean scores, which are presented in *Table 2*. The mean of the perceived publication pressure lies slightly above the mean of the scale. The effort versus reward ratio is 1.15, which means that the perceived effort put into work is higher than the perceived reward received for work. Astronomers also feel a slight overcommitment to work (M=3.39). The four forms of organisational justice are generally above the mean of the scale, which indicates that astronomers tend to feel more justice than injustice when it comes to resource allocation, peer review, grant application and telescope time application. The mean of the perceived frequency of scientific misconduct lies just below the mean of the scale (M=2.99). The mean impact of the 18 different misbehaviours (listed in *Table 2*) on the validity of the findings at hand (quality criterion 1) and on the resulting paper's ability to convey the research appropriately (quality criterion 2) are around 3.3. The mean value for the impact of misbehaviour on research diversity is higher (M=3.74). However, one needs to consider that this question was only asked for the two types of misbehaviour (Item 8 and Item 9) for which it was expected that they have an impact on quality criterion 3. That choice was made in order not to burden the participants with a question that did not fit with the rest of the misbehaviour items.

| | n valid | Mean | SD |
|---|---|---|---|
| **Independent Variables** | | | |
| Publication Pressure (PPQ) | 1949 | 3.16 | 0.77 |
| Distributive Justice: Effort | 1757 | 3.68 | 0.86 |
| Distributive Justice: Reward | 1752 | 3.21 | 0.84 |
| Organisational Justice: Resource Allocation | 1852 | 3.02 | 0.8 |
| Organisational Justice: Peer Review | 1967 | 3.69 | 0.74 |
| Organisational Justice: Grant Application | 1267 | 3.14 | 0.79 |
| Organisational Justice: Telescope Time Application | 1185 | 3.36 | 0.72 |
| Overcommitment | 1755 | 3.39 | 0.84 |
| **Dependent Variables** | | | |
| Occurrence of misconduct | 1869 | 2.99 | 0.66 |
| Impact on Quality Criterion 1 | 1868 | 3.26 | 0.62 |
| Impact on Quality Criterion 2 | 1868 | 3.29 | 0.64 |
| Impact on Quality Criterion 3 | 1902 | 3.74 | 0.93 |

*Table 2: Means and standard deviations of the independent & dependent variables*

4.2 Exploratory Factor Analyses

In order to build our SEM, we first performed an EFA for the independent variables PPQ and ERI, the latter of which includes the overcommitment items. The Pearson correlation matrix for the PPQ items show considerably low correlations for 10 items (<|0.3|). When testing the Cronbach alphas (see *S2-Appendix: S2-Table1a*) for the whole construct we found that removing those items would increase the reliability of the PPQ construct. We subsequently decided to remove those items, resulting in 12 remaining items. The Cronbach alpha for the remaining PPQ construct is 0.871 (see *S2-Table1b*), which is considered as good internal consistency. We henceforth used this cleaned PPQ for any further analysis. The CATPCA resulted in 3 factors, which we classified as F1) Extent & Consequences on one's own conduct of research, F2) Impact on Relationship with colleagues and F3) Suspected consequences on science (see *S2-Table1c*).

The EFA for the ERI construct resulted in 5 factors (see *S2-Appendix: S2-Table2a*): One factor representing the perceived effort put into the work (F3), one factor representing the perceived overcommitment to work (F5) and three factors representing the perception of being rewarded for one's work (F1: Job situation, F2: Salary, F4: Receiving praise/ respect). The Cronbach alphas are 0.691 for the effort construct, 0.780 for the overcommitment construct and 0.805 for the combined reward construct (*S2-Table2b-d*), which is considered acceptable for the former two and good for the last one.

4.3 Confirmatory Factor Analyses

We subsequently ran CFAs for all independent variables, The results are presented in *S2-Appendix: S2-Table3*. This table includes the model fit indices CFI and TLI, where >0.9 indicates a good fit for both and RMSEA, where <0.05 denotes a good fit. In addition, the fourth column includes the Chi-square values for the difference between the models with and without accounting for significant covariation between indicators measuring the respective independent variable. All independent variable constructs show a good fit according to CFI and TLI. As for RMSEA the fit is good for ERI and acceptable for the other constructs. For each independent variable, the model is unsurprisingly better when significant covariations between indicators are taken into account.

## 4.4 Structural Equation Model

This section presents the statistically significant main effects from the regression analysis of the whole structural equation model (*Figure 1*; N=520 after listwise exclusion). The SEM fit is acceptable with a CFI of 0.801, a TLI of 0.790 and a RMSEA of 0.043 90% CI(0.042, 0.044). For the sake of readability, we split the output by the independent and dependent variables, resulting in five different tables (*Table 3a* to *Table 3e*[5]; as a reminder, reference categories for the control variables are: gender: female/non-binary; academic position: full professor; primary employer: Non-Academic; institute location: Global South; number of papers published: 1–5).

*Table 3a* presents the main effects of the control variables regressed onto perceived publication pressure. Being male as opposed to female/ non-binary decreases the chance to perceive publication pressure by 0.172 points. Astronomers occupying a position other than associate professor tend to feel more publication pressure than a full professor. Whether one works at an academic institution or not does not have a significant effect on publication pressure. However, working at an institution located in the Global North are less likely to feel publication pressure by 0.33 points. As for number of published papers, the effects for most categories are not significant, apart from having published 21-30, 61-70 or more than 100 papers, which all decrease the likelihood of perceiving publication pressure as compared to having published between one and five papers.

| IV: Publication Pressure | Estimate | Std.Err | z-value | P(>\|z\|) | Std.lv | Std.all |
|---|---|---|---|---|---|---|
| Intercept: | 3.159 | | | | | |
| Gender: Male | -0.172 | 0.071 | -2.435 | 0.015* | -0.256 | -0.114 |
| Position: PhD | 0.423 | 0.176 | 2.404 | 0.016* | 0.629 | 0.115 |
| Position: Postdoc | 0.499 | 0.091 | 5.452 | < 0.001* | 0.741 | 0.307 |
| Position: Assistant Prof. | 0.397 | 0.108 | 3.679 | < 0.001* | 0.589 | 0.186 |
| Position: Associate Prof. | 0.05 | 0.09 | 0.553 | 0.58 | 0.074 | 0.028 |
| Position: Other | 0.243 | 0.105 | 2.325 | 0.02* | 0.361 | 0.123 |
| Primary Employer: Academic | 0.057 | 0.124 | 0.463 | 0.643 | 0.085 | 0.022 |
| Location: Global North | -0.33 | 0.094 | -3.517 | < 0.001* | -0.491 | -0.166 |
| Papers published: 6-10 | -0.176 | 0.113 | -1.566 | 0.117 | -0.262 | -0.084 |
| Papers published: 11-20 | 0.092 | 0.1 | 0.919 | 0.358 | 0.136 | 0.052 |
| Papers published: 21-30 | -0.284 | 0.13 | -2.176 | 0.03* | -0.422 | -0.115 |
| Papers published: 31-40 | -0.049 | 0.123 | -0.399 | 0.69 | -0.073 | -0.021 |

[5] As a reminder, reference categories for the control variables are: gender: female/ non-binary; academic position: full professor; institute location: global South; Number of papers published: 1-5.

| | | | | | | |
|---|---|---|---|---|---|---|
| Papers published: 41-50 | -0.116 | 0.123 | -0.943 | 0.346 | -0.172 | -0.05 |
| Papers published: 51-60 | -0.081 | 0.164 | -0.496 | 0.62 | -0.121 | -0.024 |
| Papers published: 61-70 | -0.335 | 0.168 | -1.991 | 0.046* | -0.498 | -0.098 |
| Papers published: 71-80 | -0.176 | 0.17 | -1.034 | 0.301 | -0.261 | -0.05 |
| Papers published: 81-90 | -0.304 | 0.243 | -1.254 | 0.21 | -0.452 | -0.059 |
| Papers published: 91-100 | -0.1 | 0.183 | -0.549 | 0.583 | -0.149 | -0.027 |
| Papers published: >100 | -0.338 | 0.137 | -2.464 | 0.014* | -0.503 | -0.133 |

*Table 3a: Extracted output for the independent variable "Publication Pressure" from the regression analysis of our SEM Model.*

As for the distributive justice factors reward and effort (*Table 3b*), astronomers who perceive publication pressure feel less rewarded (by 0.381 points) for the work they do, while at the same time they feel that they put more effort (by 0.502 points) into their work than astronomers who feel less publication pressure. Males tend to feel like they need to put less effort into their work than females or non-binaries (by 0.256 points). Postdocs tend to feel less rewarded for their work by 0.315 points, while at the same time also putting less effort than full professors by 0.263 points. Associate professors also feel that they are less rewarded as compared to full professors by 0.318 points, but show no significant effect on the effort factor. Neither being an academic astronomer, nor being employed in the global North makes a difference in the reward and effort factors as compared to the opposite. Astronomers who have published 11-20, 31-40 or 61-70 papers feel more rewarded for their job than those who have published one to five papers. Those who have published more than 100 papers are more likely to feel that they put a lot of effort into their work than the reference category (1-5 papers).

| **IV: Reward (ERI)** | **Estimate** | **Std.Err** | **z-value** | **P(>\|z\|)** | **Std.lv** | **Std.all** |
|---|---|---|---|---|---|---|
| Intercept: | 3.208 | | | | | |
| PPQ | -0.381 | 0.072 | -5.31 | < 0.001* | -0.358 | -0.358 |
| Gender: Male | 0.069 | 0.078 | 0.887 | 0.375 | 0.097 | 0.043 |
| Position: PhD | -0.151 | 0.194 | -0.779 | 0.436 | -0.211 | -0.039 |
| Position: Postdoc | -0.315 | 0.102 | -3.079 | 0.002* | -0.44 | -0.182 |
| Position: Assistant Prof. | -0.185 | 0.119 | -1.55 | 0.121 | -0.258 | -0.081 |
| Position: Associate Prof. | -0.318 | 0.101 | -3.142 | 0.002* | -0.445 | -0.168 |
| Position: Other | -0.107 | 0.115 | -0.931 | 0.352 | -0.15 | -0.051 |
| Primary Employer: Academic | 0.025 | 0.136 | 0.184 | 0.854 | 0.035 | 0.009 |
| Location: Global North | -0.113 | 0.104 | -1.093 | 0.274 | -0.158 | -0.054 |
| Papers published: 6-10 | 0.204 | 0.125 | 1.64 | 0.101 | 0.286 | 0.092 |
| Papers published: 11-20 | 0.238 | 0.111 | 2.148 | 0.032* | 0.333 | 0.128 |
| Papers published: 21-30 | 0.12 | 0.144 | 0.837 | 0.403 | 0.168 | 0.046 |
| Papers published: 31-40 | 0.353 | 0.138 | 2.568 | 0.01* | 0.494 | 0.144 |
| Papers published: 41-50 | 0.134 | 0.135 | 0.991 | 0.322 | 0.187 | 0.054 |
| Papers published: 51-60 | 0.198 | 0.181 | 1.095 | 0.274 | 0.277 | 0.056 |
| Papers published: 61-70 | 0.473 | 0.188 | 2.521 | 0.012* | 0.661 | 0.13 |
| Papers published: 71-80 | 0.25 | 0.188 | 1.333 | 0.183 | 0.35 | 0.067 |
| Papers published: 81-90 | 0.5 | 0.269 | 1.857 | 0.063 | 0.699 | 0.091 |
| Papers published: 91-100 | 0.206 | 0.202 | 1.019 | 0.308 | 0.288 | 0.051 |
| Papers published: >100 | 0.275 | 0.152 | 1.807 | 0.071 | 0.385 | 0.101 |
| **IV: Effort (ERI)** | | | | | | |
| Intercept: | 3.556 | | | | | |
| PPQ | 0.502 | 0.08 | 6.288 | < 0.001* | 0.44 | 0.44 |
| Gender: Male | -0.256 | 0.088 | -2.93 | 0.003* | -0.334 | -0.149 |
| Position: PhD | -0.217 | 0.217 | -1.001 | 0.317 | -0.283 | -0.052 |
| Position: Postdoc | -0.263 | 0.113 | -2.329 | 0.02* | -0.342 | -0.142 |
| Position: Assistant Prof. | -0.118 | 0.133 | -0.885 | 0.376 | -0.153 | -0.048 |
| Position: Associate Prof. | -0.009 | 0.111 | -0.079 | 0.937 | -0.011 | -0.004 |
| Position: Other | -0.274 | 0.129 | -2.125 | 0.034* | -0.358 | -0.121 |
| Primary Employer: Academic | -0.029 | 0.153 | -0.192 | 0.847 | -0.038 | -0.01 |
| Location: Global North | 0.156 | 0.116 | 1.346 | 0.178 | 0.203 | 0.069 |
| Papers published: 6-10 | -0.033 | 0.139 | -0.239 | 0.811 | -0.043 | -0.014 |
| Papers published: 11-20 | -0.123 | 0.123 | -0.998 | 0.318 | -0.16 | -0.062 |
| Papers published: 21-30 | 0.182 | 0.161 | 1.133 | 0.257 | 0.237 | 0.065 |
| Papers published: 31-40 | -0.081 | 0.152 | -0.533 | 0.594 | -0.106 | -0.031 |
| Papers published: 41-50 | -0.073 | 0.151 | -0.481 | 0.631 | -0.095 | -0.027 |
| Papers published: 51-60 | -0.2 | 0.202 | -0.992 | 0.321 | -0.261 | -0.053 |
| Papers published: 61-70 | 0.248 | 0.208 | 1.194 | 0.233 | 0.323 | 0.064 |
| Papers published: 71-80 | 0.034 | 0.209 | 0.163 | 0.87 | 0.045 | 0.009 |
| Papers published: 81-90 | 0.081 | 0.299 | 0.272 | 0.786 | 0.106 | 0.014 |
| Papers published: 91-100 | 0.334 | 0.226 | 1.479 | 0.139 | 0.435 | 0.077 |
| Papers published: >100 | 0.408 | 0.17 | 2.401 | 0.016* | 0.532 | 0.14 |

*Table 3b: Extracted output for the two forms of the independent variable "Distributive Justice" from the regression analysis of our SEM Model.*

The perception of all four kinds of organisational justice measured depends on the perceived publication pressure (*Table 3c*). The feeling of being treated fairly decreases with increasing publication pressure for all four latent variables (with parameter estimates between 0.273 and 0.475). PhDs and Postdocs feel treated more fairly in terms of resource allocation and peer review than full professors. PhDs also feel more justice when it comes to grant application processes. Being academic or not does not have a significant effect on any of the four organisational justice perceptions. Being employed in the global North decreases the likelihood of perceiving fairness in terms of peer review and grant application as compared to those employed in the global South. Astronomers who have published 11-20 or more than 100 papers feel treated more fairly in terms of resource allocation than those who have only published 1-5 papers. Those who have published 11-20, 31-40 or more than 100 papers feel more organisational justice in terms of peer review and grant application than the reference category (1-5 papers). Additionally, having published more than 90 papers increases the likelihood of feeling fairness in peer review processes. As for organisational justice in terms of telescope time applications, only having 11-20 as compared to 1-5 publications increases the perception of fairness significantly.

| **IV: Organisational Justice (Resource Allocation)** | **Estimate** | **Std.Err** | **z-value** | **P(>\|z\|)** | **Std.lv** | **Std.all** |
|---|---|---|---|---|---|---|
| Intercept: | 3.02 | | | | | |
| PPQ | -0.273 | 0.057 | -4.754 | < 0.001* | -0.298 | -0.298 |
| Gender: Male | 0.106 | 0.063 | 1.7 | 0.089 | 0.173 | 0.077 |
| Position: PhD | 0.389 | 0.157 | 2.475 | 0.013* | 0.632 | 0.116 |
| Position: Postdoc | 0.212 | 0.082 | 2.594 | 0.009* | 0.344 | 0.142 |
| Position: Assistant Prof. | -0.001 | 0.095 | -0.013 | 0.99 | -0.002 | -0.001 |
| Position: Associate Prof. | -0.108 | 0.08 | -1.362 | 0.173 | -0.176 | -0.067 |
| Position: Other | 0.086 | 0.092 | 0.933 | 0.351 | 0.139 | 0.047 |
| Primary Employer: Academic | 0.039 | 0.109 | 0.357 | 0.721 | 0.063 | 0.017 |
| Location: Global North | 0.069 | 0.083 | 0.83 | 0.406 | 0.111 | 0.038 |
| Papers published: 6-10 | 0.071 | 0.099 | 0.717 | 0.474 | 0.115 | 0.037 |
| Papers published: 11-20 | 0.206 | 0.089 | 2.308 | 0.021* | 0.334 | 0.128 |
| Papers published: 21-30 | 0.058 | 0.115 | 0.503 | 0.615 | 0.094 | 0.025 |
| Papers published: 31-40 | 0.083 | 0.108 | 0.766 | 0.444 | 0.135 | 0.039 |
| Papers published: 41-50 | 0.028 | 0.108 | 0.256 | 0.798 | 0.045 | 0.013 |
| Papers published: 51-60 | 0.023 | 0.144 | 0.16 | 0.873 | 0.037 | 0.008 |
| Papers published: 61-70 | 0.04 | 0.148 | 0.273 | 0.785 | 0.066 | 0.013 |
| Papers published: 71-80 | 0.107 | 0.15 | 0.718 | 0.473 | 0.174 | 0.034 |
| Papers published: 81-90 | 0.127 | 0.214 | 0.595 | 0.552 | 0.206 | 0.027 |
| Papers published: 91-100 | 0.119 | 0.161 | 0.741 | 0.459 | 0.194 | 0.034 |
| Papers published: >100 | 0.279 | 0.123 | 2.279 | 0.023* | 0.453 | 0.119 |
| **IV: Organisational Justice (Peer Review)** | | | | | | |
| Intercept: | 3.692 | | | | | |
| PPQ | -0.42 | 0.074 | -5.688 | < 0.001* | -0.343 | -0.343 |
| Gender: Male | 0.015 | 0.082 | 0.184 | 0.854 | 0.018 | 0.008 |
| Position: PhD | 0.563 | 0.206 | 2.741 | 0.006* | 0.683 | 0.125 |
| Position: Postdoc | 0.241 | 0.106 | 2.265 | 0.024* | 0.292 | 0.121 |
| Position: Assistant Prof. | 0.197 | 0.125 | 1.572 | 0.116 | 0.239 | 0.075 |
| Position: Associate Prof. | 0.066 | 0.105 | 0.63 | 0.529 | 0.08 | 0.03 |
| Position: Other | 0.156 | 0.122 | 1.28 | 0.201 | 0.189 | 0.064 |
| Primary Employer: Academic | 0.153 | 0.144 | 1.065 | 0.287 | 0.186 | 0.049 |
| Location: Global North | -0.44 | 0.11 | -3.998 | < 0.001* | -0.534 | -0.181 |
| Papers published: 6-10 | 0.21 | 0.131 | 1.599 | 0.11 | 0.254 | 0.082 |
| Papers published: 11-20 | 0.403 | 0.117 | 3.442 | 0.001* | 0.488 | 0.188 |
| Papers published: 21-30 | 0.123 | 0.152 | 0.81 | 0.418 | 0.149 | 0.041 |
| Papers published: 31-40 | 0.312 | 0.144 | 2.169 | 0.03* | 0.378 | 0.11 |
| Papers published: 41-50 | 0.13 | 0.143 | 0.909 | 0.364 | 0.157 | 0.045 |
| Papers published: 51-60 | -0.078 | 0.19 | -0.409 | 0.682 | -0.094 | -0.019 |
| Papers published: 61-70 | 0.369 | 0.196 | 1.883 | 0.06 | 0.448 | 0.088 |
| Papers published: 71-80 | 0.261 | 0.198 | 1.319 | 0.187 | 0.316 | 0.061 |
| Papers published: 81-90 | 0.517 | 0.283 | 1.828 | 0.067 | 0.627 | 0.082 |
| Papers published: 91-100 | 0.64 | 0.214 | 2.99 | 0.003* | 0.776 | 0.138 |
| Papers published: >100 | 0.324 | 0.16 | 2.021 | 0.043* | 0.392 | 0.103 |
| **IV: Organisational Justice (Grant Application)** | **Estimate** | **Std.Err** | **z-value** | **P(>\|z\|)** | **Std.lv** | **Std.all** |
| Intercept: | 3.143 | | | | | |
| PPQ | -0.475 | 0.082 | -5.828 | < 0.001* | -0.357 | -0.357 |
| Gender: Male | 0.066 | 0.09 | 0.737 | 0.461 | 0.074 | 0.033 |
| Position: PhD | 0.71 | 0.225 | 3.157 | 0.002* | 0.794 | 0.145 |
| Position: Postdoc | 0.192 | 0.116 | 1.659 | 0.097 | 0.215 | 0.089 |
| Position: Assistant Prof. | 0.156 | 0.137 | 1.137 | 0.255 | 0.174 | 0.055 |
| Position: Associate Prof. | -0.174 | 0.115 | -1.516 | 0.13 | -0.194 | -0.073 |
| Position: Other | 0.168 | 0.133 | 1.267 | 0.205 | 0.188 | 0.064 |
| Primary Employer: Academic | 0.139 | 0.157 | 0.885 | 0.376 | 0.155 | 0.041 |
| Location: Global North | -0.246 | 0.119 | -2.058 | 0.04* | -0.275 | -0.093 |
| Papers published: 6-10 | 0.033 | 0.143 | 0.231 | 0.817 | 0.037 | 0.012 |
| Papers published: 11-20 | 0.319 | 0.127 | 2.503 | 0.012* | 0.356 | 0.137 |
| Papers published: 21-30 | 0.122 | 0.166 | 0.738 | 0.46 | 0.137 | 0.037 |
| Papers published: 31-40 | 0.314 | 0.157 | 2.003 | 0.045* | 0.351 | 0.103 |
| Papers published: 41-50 | -0.003 | 0.156 | -0.02 | 0.984 | -0.004 | -0.001 |
| Papers published: 51-60 | 0.074 | 0.208 | 0.358 | 0.72 | 0.083 | 0.017 |
| Papers published: 61-70 | 0.325 | 0.214 | 1.521 | 0.128 | 0.364 | 0.072 |
| Papers published: 71-80 | 0.33 | 0.216 | 1.526 | 0.127 | 0.368 | 0.071 |
| Papers published: 81-90 | 0.26 | 0.308 | 0.843 | 0.399 | 0.291 | 0.038 |
| Papers published: 91-100 | 0.057 | 0.232 | 0.246 | 0.806 | 0.064 | 0.011 |
| Papers published: >100 | 0.393 | 0.175 | 2.245 | 0.025* | 0.439 | 0.116 |

| **IV: Organisational Justice (Telescope Time Application)** | | | | | | |
|---|---|---|---|---|---|---|
| Intercept: | 3.356 | | | | | |
| PPQ | -0.299 | 0.07 | -4.292 | < 0.001* | -0.256 | -0.256 |
| Gender: Male | 0.114 | 0.081 | 1.407 | 0.159 | 0.145 | 0.065 |
| Position: PhD | 0.12 | 0.201 | 0.6 | 0.549 | 0.154 | 0.028 |
| Position: Postdoc | 0.074 | 0.104 | 0.712 | 0.477 | 0.095 | 0.039 |
| Position: Assistant Prof. | -0.016 | 0.123 | -0.134 | 0.894 | -0.021 | -0.007 |
| Position: Associate Prof. | 0.008 | 0.103 | 0.08 | 0.936 | 0.01 | 0.004 |
| Position: Other | 0.101 | 0.119 | 0.847 | 0.397 | 0.129 | 0.044 |
| Primary Employer: Academic | -0.04 | 0.141 | -0.283 | 0.777 | -0.051 | -0.013 |
| Location: Global North | -0.186 | 0.107 | -1.737 | 0.082 | -0.238 | -0.081 |
| Papers published: 6-10 | -0.009 | 0.128 | -0.067 | 0.946 | -0.011 | -0.004 |
| Papers published: 11-20 | 0.251 | 0.114 | 2.192 | 0.028* | 0.32 | 0.123 |
| Papers published: 21-30 | 0.231 | 0.149 | 1.548 | 0.122 | 0.295 | 0.08 |
| Papers published: 31-40 | 0.134 | 0.141 | 0.949 | 0.342 | 0.171 | 0.05 |
| Papers published: 41-50 | -0.043 | 0.14 | -0.308 | 0.758 | -0.055 | -0.016 |
| Papers published: 51-60 | -0.086 | 0.187 | -0.46 | 0.646 | -0.11 | -0.022 |
| Papers published: 61-70 | 0.237 | 0.192 | 1.231 | 0.218 | 0.302 | 0.06 |
| Papers published: 71-80 | 0.212 | 0.194 | 1.09 | 0.276 | 0.27 | 0.052 |
| Papers published: 81-90 | 0.386 | 0.278 | 1.389 | 0.165 | 0.493 | 0.064 |
| Papers published: 91-100 | 0.004 | 0.209 | 0.021 | 0.983 | 0.006 | 0.001 |
| Papers published: >100 | 0.098 | 0.157 | 0.623 | 0.534 | 0.125 | 0.033 |

*Table 3c: Extracted output for the four forms of the independent variable "Organisational Justice" from the regression analysis of our SEM Model.*

The feeling of being overcommitted to work unsurprisingly significantly increases with increasing perceived effort put into work by 0.464 points (*Table 3d*). Associate professors tend to feel less overcommitted to their work compared to full professors by 0.166 points.

| **IV: Overcommitment (ERI)** | **Estimate** | **Std.Err** | **z-value** | **P(>\|z\|)** | **Std.lv** | **Std.all** |
|---|---|---|---|---|---|---|
| Intercept: | 3.433 | | | | | |
| Publication Pressure | 0.113 | 0.06 | 1.866 | 0.062 | 0.13 | 0.13 |
| Reward (ERI) | -0.083 | 0.075 | -1.103 | 0.27 | -0.102 | -0.102 |

| Effort (RI) | 0.464 | 0.061 | 7.658 | < 0.001* | 0.609 | 0.609 |
|---|---|---|---|---|---|---|
| Organisational Justice (RA) | -0.024 | 0.073 | -0.323 | 0.747 | -0.025 | -0.025 |
| Organisational Justice (PR) | 0.058 | 0.038 | 1.527 | 0.127 | 0.082 | 0.082 |
| Organisational Justice (GA) | 0.005 | 0.037 | 0.147 | 0.883 | 0.008 | 0.008 |
| Organisational Justice (TA) | -0.066 | 0.041 | -1.595 | 0.111 | -0.088 | -0.088 |
| Gender: Male | 0.011 | 0.057 | 0.195 | 0.845 | 0.019 | 0.009 |
| Position: PhD | 0.137 | 0.146 | 0.934 | 0.35 | 0.234 | 0.043 |
| Position: Postdoc | 0.032 | 0.082 | 0.388 | 0.698 | 0.054 | 0.023 |
| Position: Assistant Prof. | -0.071 | 0.086 | -0.815 | 0.415 | -0.121 | -0.038 |
| Position: Associate Prof. | -0.166 | 0.074 | -2.239 | 0.025* | -0.284 | -0.107 |
| Position: Other | -0.021 | 0.085 | -0.245 | 0.807 | -0.035 | -0.012 |
| Primary Employer: Academic | -0.065 | 0.098 | -0.67 | 0.503 | -0.112 | -0.029 |
| Location: Global North | 0.042 | 0.077 | 0.539 | 0.59 | 0.071 | 0.024 |
| Papers published: 6-10 | 0.032 | 0.09 | 0.358 | 0.721 | 0.055 | 0.018 |
| Papers published: 11-20 | -0.011 | 0.08 | -0.136 | 0.892 | -0.019 | -0.007 |
| Papers published: 21-30 | -0.027 | 0.103 | -0.263 | 0.793 | -0.046 | -0.013 |
| Papers published: 31-40 | 0.021 | 0.1 | 0.209 | 0.834 | 0.036 | 0.01 |
| Papers published: 41-50 | 0.095 | 0.097 | 0.975 | 0.33 | 0.162 | 0.047 |
| Papers published: 51-60 | -0.006 | 0.13 | -0.048 | 0.961 | -0.011 | -0.002 |
| Papers published: 61-70 | 0.158 | 0.137 | 1.152 | 0.249 | 0.271 | 0.053 |
| Papers published: 71-80 | -0.076 | 0.134 | -0.567 | 0.571 | -0.13 | -0.025 |
| Papers published: 81-90 | -0.223 | 0.194 | -1.146 | 0.252 | -0.381 | -0.05 |
| Papers published: 91-100 | 0.102 | 0.147 | 0.693 | 0.488 | 0.174 | 0.031 |
| Papers published: >100 | -0.153 | 0.111 | -1.372 | 0.17 | -0.261 | -0.069 |

*Table 3d: Extracted output for the independent variable "Overcommitment" from the regression analysis of our SEM Model.*

Finally, we turn to the results of our dependent variable: perception of how often misconduct occurs in astronomy (*Table 3e*). The parameter estimate of the main effect of publication pressure on perception of misconduct frequency is 0.375, which means that increasing perceived publication pressure leads to increased perception of misconduct. Perceived fairness of telescope time application processes also has a significant effect on the perception of misconduct; a decreasing feeling of fairness increases the perception of misconduct by 0.113 points. Being employed in the global North also increases the perception of misconduct by 0.212 points. In addition to the main effects on perceived frequency of misconduct, our model calculates the mediated effects. Let us first attend the results of the effects of the control variables as mediated by publication pressure. Being male as compared to female/ non-binary reduces the perception of misconduct by 0.064 points. Any other position than associate professor increases the chance to perceive misconduct as compared to full professor. Employment at an institution in the global North decreases the perception of misconduct by 0.124 points, mediated through the perception publication pressure. Having published 21-30 or more than 100 papers also decreases the likelihood of perceiving misconduct.

| **DV: Frequency of Misbehaviour Occurrence** | **Estimate** | **Std.Err** | **z-value** | **P(>\|z\|)** | **Std.lv** | **Std.all** |
|---|---|---|---|---|---|---|
| Intercept: | 2.992 | | | | | |
| Publication Pressure | 0.375 | 0.067 | 5.616 | < 0.001* | 0.438 | 0.438 |
| Reward (ERI) | -0.135 | 0.073 | -1.838 | 0.066 | -0.168 | -0.168 |
| Effort (RI) | -0.02 | 0.058 | -0.337 | 0.736 | -0.026 | -0.026 |
| Organisational Justice (RA) | 0.09 | 0.069 | 1.303 | 0.192 | 0.097 | 0.097 |
| Organisational Justice (PR) | -0.054 | 0.036 | -1.493 | 0.135 | -0.077 | -0.077 |
| Organisational Justice (GA) | -0.004 | 0.035 | -0.122 | 0.903 | -0.007 | -0.007 |
| Organisational Justice (TA) | -0.113 | 0.04 | -2.835 | 0.005* | -0.154 | -0.154 |
| Overcommitment (ERI) | 0.074 | 0.073 | 1.006 | 0.314 | 0.075 | 0.075 |
| Gender: Male | -0.055 | 0.054 | -1.032 | 0.302 | -0.096 | -0.043 |
| Position: PhD | -0.065 | 0.138 | -0.47 | 0.638 | -0.113 | -0.021 |
| Position: Postdoc | 0.005 | 0.077 | 0.065 | 0.948 | 0.009 | 0.004 |
| Position: Assistant Prof. | 0.054 | 0.082 | 0.664 | 0.507 | 0.094 | 0.03 |
| Position: Associate Prof. | -0.02 | 0.07 | -0.288 | 0.774 | -0.035 | -0.013 |
| Position: Other | -0.049 | 0.08 | -0.619 | 0.536 | -0.086 | -0.029 |
| Primary Employer: Academic | -0.179 | 0.093 | -1.933 | 0.053 | -0.311 | -0.082 |
| Location: Global North | 0.212 | 0.074 | 2.851 | 0.004* | 0.368 | 0.125 |
| Papers published: 6-10 | -0.009 | 0.084 | -0.112 | 0.91 | -0.016 | -0.005 |
| Papers published: 11-20 | 0.01 | 0.076 | 0.134 | 0.893 | 0.018 | 0.007 |
| Papers published: 21-30 | 0.171 | 0.098 | 1.753 | 0.08 | 0.298 | 0.081 |
| Papers published: 31-40 | 0.027 | 0.094 | 0.284 | 0.776 | 0.046 | 0.014 |
| Papers published: 41-50 | 0.093 | 0.092 | 1.01 | 0.313 | 0.161 | 0.047 |
| Papers published: 51-60 | 0.075 | 0.122 | 0.617 | 0.537 | 0.131 | 0.026 |
| Papers published: 61-70 | 0.131 | 0.13 | 1.007 | 0.314 | 0.227 | 0.045 |
| Papers published: 71-80 | 0.08 | 0.127 | 0.631 | 0.528 | 0.139 | 0.027 |
| Papers published: 81-90 | 0.029 | 0.183 | 0.16 | 0.873 | 0.051 | 0.007 |
| Papers published: 91-100 | -0.104 | 0.138 | -0.754 | 0.451 | -0.181 | -0.032 |
| Papers published: >100 | 0.037 | 0.105 | 0.354 | 0.723 | 0.065 | 0.017 |

*Table 3e: Extracted output for the dependent variable "Frequency of Misbehaviour Occurrence" from the regression analysis of our SEM Model.*

### 4.5 Perceived impact on research quality

Lastly, we analysed the perceived impact of the 18 types of misbehaviour on the three aspects of research quality: the validity of the findings at hand (quality criterion 1; QC1), the resulting paper's ability to convey the research appropriately (quality criterion 2; QC2) and the impact on research diversity (quality criterion 3; QC3). Following Bouter et al. (2016) and Haven et al. (under review), we calculated the perceived impact as the product score of the means of perceived frequency and impact on the respective quality criterion (means are listed in *S3-Appendix: S3-Table1*). The higher the resulting number, the higher the harm on research. *S3-Table2* lists the results of perceived impact scores in descending order. In addition, *Figure 2* visualises the perceived impact on research quality in four quadrants, indicating high frequency and high impact (Q1), low frequency and low impact (Q3), low frequency and high impact (Q2) and high frequency and low impact (Q4). Item 8 ("Propose study questions solely because they are considered a 'hot' topic") and Item 9 ("Not considering a study question because it isn't considered a 'hot' topic, even though it could be important for astronomy") show the highest perceived impact on research quality (14.33 and 12.89, respectively). This impact is related to QC3 ("impact on research diversity") and can be found in Q1. While the high impact of these types of misbehaviour on QC3 (M=3.7, M=3.79 for Item 8 & 9, respectively) may well be expected, the comparatively high frequencies (M=3.88, M=3.4 for Item 8 & 9, respectively) are an interesting result. Item 18 (“Biased interpretation of data that distorts results”) follows, with perceived impact amounting to 12.47 for QC1 and 12.17 for QC2. As we can see from *Figure 2*, these high values for perceived impact stem mostly from the comparatively high impacts on the two quality criteria (M=4.04 & M=3.95), rather than a high frequency (M=3.08), which is just above the mean (2.99). The occurrence of Item 13 (“Data fabrication and/ or falsification”) also has a high impact on both quality criteria (M=4.17 for QC1, M=4.01 for QC2). Due to the comparative low frequency of Item 13 (M=1.95), the perceived impact of this type of misbehaviour corresponds to rank 30 (QC1) and 31 (QC2) out of 38. Item 13 can be found in Q2. Item 10 (“Giving authorship credit to someone who has not contributed substantively to a manuscript”) which is ranked lower than Item 13 (32 for QC2 and 31 for QC1) can be interpreted as the opposite of Item 13 and hence be found in Q4; Item 10 has a high occurrence (M=3.73), but low impact on quality criteria (M=1.97 for QC1 and M=2.06 for QC2). By contrast, “Denying authorship credit to someone who has contributed substantively to a manuscript” (Item 11) is located in Q3, since it doesn’t occur very often (M=2.05) and when it occurs it also has a comparatively low impact on QC1 (M=2.53) and QC2 (M=2.67). The perceived impact of this type of misbehaviour is also the lowest ranked.

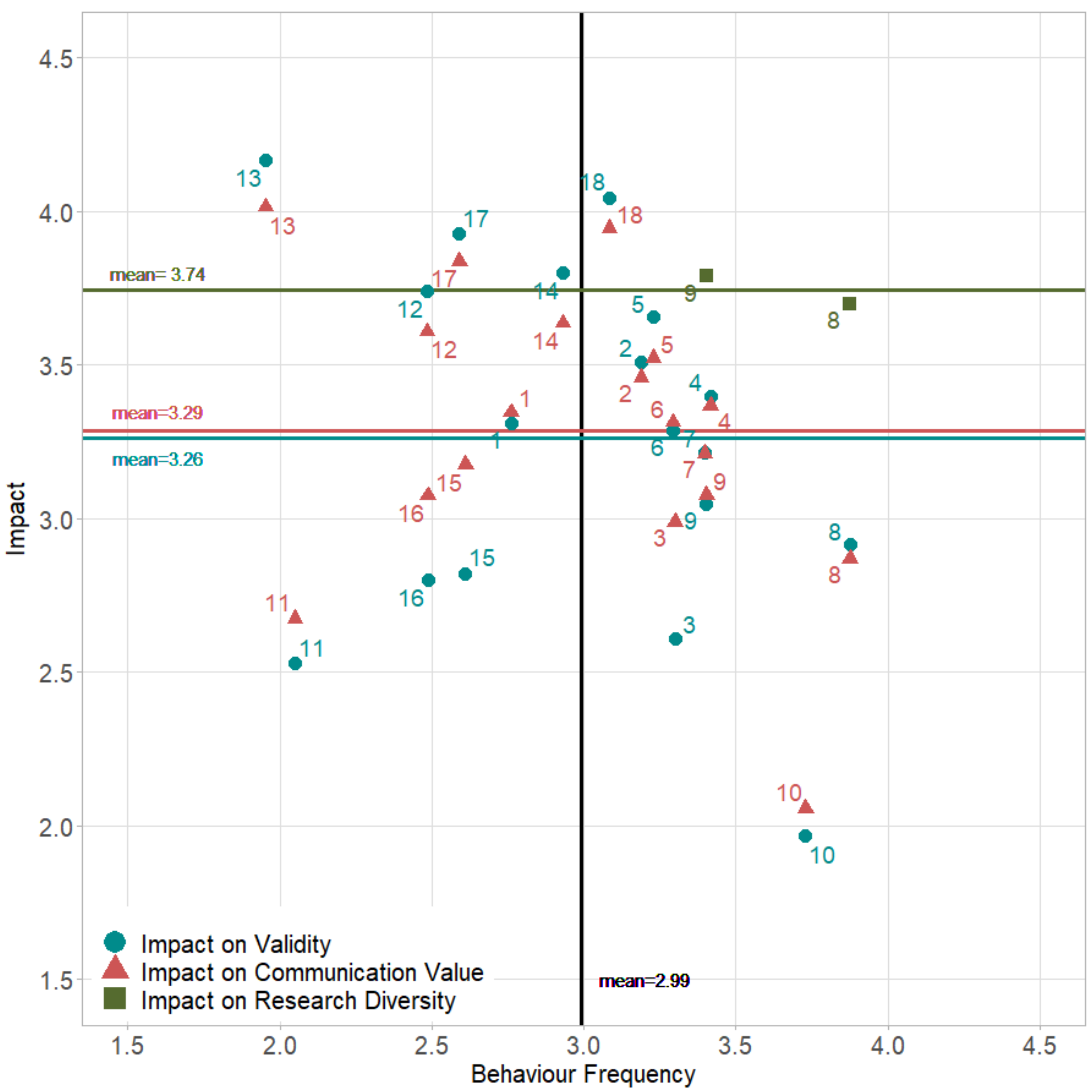

*Figure 2: Impact on the three aspects of research quality for each type of misbehaviour versus the frequency of occurrence of each type of misbehaviour. Circles denote QC1, triangles refer to QC2 and the square to QC3. The four quadrants indicate high frequency and high impact (Q1), low frequency and low impact (Q3), low frequency and high impact (Q2) and high frequency and low impact (Q4).*

## 5. Discussion

Building on previous quantitative research on scientific misconduct (Bouter et al, 2016; Martinson et al., 2005, 2006, 2009, 2010 & 2016; Haven, 2021) and qualitative research on deviant behaviour in astronomy (Heuritsch, 2019; Heuritsch, 2021), we built a structural equation model relating role-associated factors, such as academic position and location of employment with environmental factors, such as perceived publication pressure and distributive & organisational justice and our dependent variables; scientific misconduct and research quality. We found that the location of the institution where an astronomer is employed in terms of the Global North versus Global South makes up about 8% of the variance of observed misconduct, considering both effects; the direct one and the suppressed one through publication pressure as a mediator. Perceived organisational justice in terms of telescope time application processes explains 2% and perceived publication pressure explains nearly 19% of the variance of observed misconduct.

In addition of publication pressure having a direct effect on scientific misconduct (cf. Haven et al., under review), we worked from the assumption that perceived publication pressure influences the perception of distributive and organisational justice, which our results confirm. An astronomer who perceives publication pressure is more likely to perceive less reward from their work, less organisational justice (in terms of resource allocation, peer review, grant application and telescope time application) and the need to put more effort into work. Hence, publication pressure is indeed a key factor to determine how research culture and integrity in astronomy is perceived.

Publication pressure, in turn, is more likely to be perceived by astronomers with academic ranks below a full professor (cf. Miller et al., 2011; Tijdink et al., 2014b). Interestingly, astronomers employed at institutions in the global North feel less publication pressure, despite observing more misconduct. Hence, there is some difference between institutions in the global North as compared to the global South, which makes astronomers perceive more publication pressure (which in turn increases the likelihood of observing misconduct), yet at the same time suppresses the perception of scientific misconduct. There is a tendency to perceive less publication pressure, when one has published more than 1-5 first or co-authored papers in the last 5 years. However, since there is only a statistical significance for 3 out of 11 categories, this requires further investigation. Males do not only perceive less publication pressure than females/non-binaries, but also feel like they need to put less effort into their work, which is consistent with previous research showing that working conditions are harder for females than males[6].

Early career researchers perceive more organisational justice in terms of resource allocation, peer review and grant allocation than full professors. This may be because early career

---

[6] For a good overview of such studies refer to: https://theconversation.com/amp/looking-at-the-stars-or-falling-by-the-wayside-how-astronomy-is-failing-female-scientists-159139 (18th June 2021)

researchers may still have a positive opinion of organisational processes, whereas more experience may lead to more occasions of unfairness (cf. Heuritsch, 2021).

The frequency of observed misbehaviour shows that the severe types of misbehaviour (FFP type) such as data fabrication & falsification (Item 13), concealing results (Item 17) and forms of plagiarism (Items 15 & 16) occur less often than the QRPs, as expected (Martinson et al., 2005; Haven, 2021). Among the most frequently occurring QRPs are making the topic selection dependent on whether a topic is or isn't "hot" (Items 8 & 9), questionable authorship practices (Item 10) and insufficient supervision (Item 4). Our findings agree with (Bouter et al., 2016) who found that data fabrication & falsification (Item 13) is believed to be the biggest threat to the validity of the findings (QC1) at hand and the communication value of the resulting paper (QC2). In comparison, plagiarism (an FFP type misbehaviour) is ranked very low on the impact on QC1 and QC2 scores. This makes sense, since as Haven et al. (under review) point out, "plagiarism fails to connect the knowledge to its proper origin, but it need not distort scientific knowledge per se", whereas falsification & fabrication do. Our perceived impact ranking (the product of the mean scores of perceived frequency and impact on research quality) also agrees with the findings of Bouter et al. (2016). It suggests that is not outright fraud which dominates the negative impact on knowledge creation, but rather behaviour which cuts corners to publishable output (Item 8, 9 & 5). We conclude that many a little makes a mickle; the epistemic harm in research in astronomy done by QRPs seems to be greater than that done by FFPs, which would agree with finding by Haven (2021) and Bouter et al. (2016).

## 6. Strengths/ Limitations

Surveys and data analysis come with strengths and limitations. Let us first turn to the strengths of this study. First, we sampled astronomers from all over the world. This paper gives a snapshot of the international cultural climate in astronomy and its impact on scientific misconduct and quality, while previous studies on this topic mainly focused on universities in the US or the Netherlands (e.g. Wells et al., 2014; Haven et al., 2019a). The second strong point of this survey is that the types of misconduct we chose are not only based on previous literature on this topic (e.g. Martinson et al., 2005, 2006, 2009, 2010 & 2016; Haven, 2021), but also on qualitative research conducted on deviant behaviour in the field of astronomy (Heuritsch, 2021). As Hesselmann (2014; p.61f.) points out, "the meaning of misbehaviour is permanently shifting" and therefore, "measuring scientific misconduct quantitatively should not be first on our research agenda." We believe that the qualitative study by Heuritsch (2021) gave us solid ground to tailor the quantitative studies performed on misbehaviour to the field of astronomy. Third, our analysis is the first among the literature relating publication pressure, distributive & organisational justice with scientific misconduct, which uses structural equation modelling, allowing for an estimation of the model in its whole complexity. Fourth, it is also the first study in this set of literature, which operationalises research quality. We therefore measure the impact of scientific misconduct on scientific quality instead of implying that relationship through the concept of research integrity.

Our study also comes with several limitations. First, while our response rate was acceptable (25%), our competition rate lies at around 14%, which may be considered as relatively low, but can still be compared to that of similar web-based surveys (e.g. Haven et al., 2019a). The reason for this drop may be that the survey was considered long, as evident from some feedback from the respondents; it took 30-60 minutes to complete. Second, due to the length of the survey, we may need to consider response bias towards those who feel publication pressure and feel treated unfairly and may hence be more enthusiastic in voicing their opinion about this topic. On the one hand, this may overestimate the effects of publication pressure and organisational injustice on misconduct. On the other hand, those who left the field of Astronomy as a result of publication pressure and injustice are of course not sampled, and hence publication pressure and organisational injustice may be underestimated through survivor bias (Kurtz & Henneken, 2017). Third, respondents criticised that there was no "NA" option for the experience of misbehaviour items. While respondents didn't have to choose an answer to move forward in the survey, this may have resulted in an underestimation of occurrence of misconduct, as astronomers who may not have much experience in the field, clicked the lowest or the middle answer category, while having preferred an NA option. Because self-reports may result in underreporting misbehaviour (Hesselmann, 2014) we chose to ask about the general experience with the types of misbehaviour. Therefore, we expect that the underreporting of one's own misconduct would mitigate the overreporting of misconduct by others (Haven et al., under review). Fourth, many filter questions, such as if one has already applied for grants or telescope time, resulted in a comparative small sample (N=520) for the SEM analysis, because of list-wise exclusion. Fifth, at the time we designed the survey we had no knowledge about the revised PPQ (Haven et al.; 2019b), which we would have used instead of the PPQ and may have resulted into better construct validity, without having to adapt the construct for our own further analysis. Sixth, while it was our theoretical aim to conduct a census of all worldwide astronomers, this was practically impossible due to time, budget and resource constraints. Our three-stage sampling design aimed at completeness in getting a hold on astronomical institutions worldwide and reaching as many astronomers as possible. However, we cannot expect that our list is indeed complete, nor that our contacts have reached all astronomers from the respective institutes, nor that our sample is random. Therefore, representativeness may be limited. To improve this, further testing for item batteries may also inform weighting to adjust the sample proportions to the population proportion.

## 7. Conclusion, Implications and Outlook for Further Research

The aim of this research was to study the impact of perceptions of publication pressure and distributive & organisational justice on the observation of occurrence of scientific misconduct and the impact that certain types of misconduct have on research quality in astronomy. While we did not find statistically significant effects of perceived distributive & organisational justice of the four processes – resource allocation, peer review, grant application and telescope time application – on research misbehaviour, we strongly emphasise that publication pressure is part of research culture. As outlined by Heuritsch (2021), institutional norms define what is seen as a good researcher, and publication rate is one of the key indicators to measure the performance of an astronomer. Arguably, playing the indicator game is an innovative path to success, so we worked from the assumption that research (mis-)behaviour is reflexive, insofar that it depends on how one's performance is evaluated. We found that publication pressure explains 19% of the variance of occurrence of misconduct and between 7 and 13% of the variance of the perception of distributive & organisational justice as well as overcommitment to work. We subsequently analysed the impact of the individual types of misbehaviour on three aspects of research quality. We agree with findings of previous studies (e.g. Haven, 2021; Bouter et al., 2016) that QRPs should not be underestimated for their epistemic harm.

We conclude that there is a need for a policy change. In the distribution of institutional rewards, grants and telescope time, less attention to metrics (such as publication rate) would foster better scientific conduct and hence research quality. Publication pressure could also be reduced by reconsidering what is considered publishable. Since, for example, negative results cannot easily be published (Heuritsch, 2021), a lot of scientific work may not be recognised as valuable research. Future studies could be devoted to exploring potentially more innovative ways of creating quality output which may count towards one's performance. This requires reflecting and working on the structural conditions that comprise the norms and culture of research in astronomy. After all, they comprise the external constituents of the *situation* of an actor, and are therefore of high relevance in the individual's actions. Future quantitative studies may complete the rational choice picture, by paying tribute to the internal component of the astronomers' research situation. For example, one could study their motivation to do research and to publish, and could relate this to the importance of publications in the field.

## Acknowledgements

First, I would like to extend my gratitude to the 3,509 astronomers who dedicated upwards of half an hour – despite publish-or-perish – to participate in this survey. Second, Thea Gronemeier and Florian Beng assisted the survey design and were a big support in data processing. Third, I appreciated the discussions with Helmut Schöller regarding the relationships between the variables and that he provided me with the Idiographic System Modelling Tool to draw the SEM. Fourth, thank you to all the prestesters; Niels Taubert, Jens Ambrasat, Andrej Dvornik, Iva Laginja, Levente Borvák, Nathalie Schwichtenberg, Theresa Velden, Richard Heidler, Rudolf Albrecht, Andreas Herdin, Alexander Fenton and Philipp Löschl. Fifth, I would like to thank Enrique Garcia Bourne for the mental support throughout, the proofreading and for assisting me with the script that helped reach out to thousands of IAU members. Finally, I am grateful to GESIS (Lebniz Institut für Sozialwissenschaften) for the scholarship that enabled me to participate in their 2019 survey methodology course, where I met Sonila Dardha – an extraordinarily competent and kind survey methodologist, whose feedback was vital for this paper.

## Conflict of Interest

The authors declare no conflict of interest.

## Funding

This study was performed in the framework of the junior research group "Reflexive Metrics", which is funded by the BMBF (German Bundesministerium für Bildung und Forschung; project number: 01PQ17002).

## References

Agnew R., (1992), "Foundation for a general strain theory of crime and delinquency", *Criminology;30*: p.47–87

Anderson, M.S., Ronning, E.A., De Vries, R., Martinson, B.C., "The Perverse Effects of Competition on Scientists' Work and Relationships", *Sci Eng Ethics (2007) 13*: p.437–461 https://doi.org/10.1007/s11948-007-9042-5

Atkinson-Grosjean, J. & Fairley, C., (2009), "Moral Economies in Science: From Ideal to Pragmatic", *Minerva*, p.147–170, https://doi.org/10.1007/s11024-009-9121-7

Bedeian A., Taylor S., Miller A. (2010), “Management science on the credibility bubble: Cardinal sins and various misdemeanors”, *Acad Manag Learn Educ*;*9(4)*: p.715–25

Bouter L.M. (2015), “Commentary: Perverse incentives or rotten apples?”, *Account Res.;22(3)*: p.148–61

Bouter, L.M., Tijdink, J., Axelsen, N., Martinson, B.C., ter Riet, G. (2016), “Ranking major and minor research misbehaviors: results from a survey among participants of four World Conferences on Research Integrity”, *Res Integr Peer Rev.;1(17)*: p.1–8

Chang & Huang (2015), “The effects of research resources on international collaboration in the astronomy community“

Coleman, J.S. (1990), “Foundations of Social Theory”, *Cambridge, Mass., and London*

Crain, L.A., Martinson, B.C., Thrush, C.R. (2013), “Relationships between the Survey of Organizational Research Climate (SORC) and self-reported research practices”, *Sci Eng Ethics.;19(3)*: p.835–50

Dahler-Larsen, P. (2014), “Constitutive Effects of Performance Indicators: Getting beyond unintended consequences”, *Public Management Review*, 16:7, p.969-986

Dahler-Larsen, P. (2019), “Quality – From Plato to Performance”, *Palgrave Macmillan*, https://doi.org/10.1007/978-3-030-10392-7

Desrosières, A. (1998), “The Politics of Large Numbers – A History of Statistical Reasoning”, *Harvard University Press*, ISBN 9780674009691

Espeland, W.N. & Vannebo B. (2008), “*Accountability, Quantification, and Law*”, Annual Review of Law and Social Science 3, p.21-43

Esser, H. (1999), “Soziologie. Spezielle Grundlagen. Band 1: Situationslogik und Handeln”, *Campus Verlag Frankfurt/New York*, https://doi.org/10.1007/s11577-001-0109-z

Fochler, M. & De Rijcke, S. (2017), “Implicated in the Indicator Game? An Experimental Debate”, *Engaging Science, Technology, and Society 3*, p.21-40

Hackett E.J (1994), “A Social Control Perspective on Scientific Misconduct”, *J Higher Educ*.;65(3): p.242–60

Halffman, W., Radder, H. (2015), “The Academic Manifesto: From an Occupied to a Public University”, *Minerva;53(2)*: p.165–87

Haven, T.L. (2021), “Towards a responsible research climate: findings from academic research in Amsterdam”, https://research.vu.nl/en/publications/towards-a-responsible-research-climate-findings-from-academic-res

Haven, T.L., Bouter, L.M., Smulders, Y.M., Tijdink, J.K. (2019a), “Perceived publication pressure in Amsterdam – survey of all disciplinary fields and academic ranks”, *PLoS One;14(6)*

Haven, T.L., Tijdink, J.K., De Goede, M.E.E., Oort, F. (2019b), “Personally perceived publication pressure - Revising the Publication Pressure Questionnaire (PPQ) by using work stress models”, *Res Integr Peer Rev.;4(7):* p.1–9, https://doi.org/10.1186/s41073-019-0066-6

Haven, T.L., Tijdink, J.K., Martinson, B.C., Bouter, L., Oort, F. (under review), “Explaining variance in perceived research misbehaviour”; in Haven (2021)

Haven, T.L., van Woudenberg, R. (in print), "Explanations of Research Misconduct, and How They Hang Together", *Accepted for publication in Journal for General Philosophy of Science;* in Haven (2021)

Heidler, R., (2011), "Cognitive and Social Structure of the Elite Collaboration Network of Astrophysics: A Case Study on Shifting Network Structures", *Minerva*, p.461–488, https://doi.org/10.1007/s11024-011-9184-0

Hesselmann, F., Wienefoet, V., & Reinhart, M. (2014), "Measuring Scientific Misconduct – Lessons from Criminology", *Publications, 2(3)*: p.61-70, https://doi.org/10.3390/publications2030061

Heuritsch, J. (2019), "Effects of metrics in research evaluation on knowledge production in astronomy A case study on Evaluation Gap and Constitutive Effects", *Proceedings of the STS Conference Graz 2019*, https://doi.org/10.3217/978-3-85125-668-0-09

Heuritsch, J. (2021), "The Evaluation Gap in Astronomy – Explained through a Rational Choice Framework", https://arxiv.org/abs/2101.03068

Kurtz, M. J. & Henneken, E. A., (2017), "Measuring Metrics - A 40-Year Longitudinal Cross-Validation of Citations, Downloads, and Peer Review in Astrophysics", *Journal of the Association for Information Science and Technology*, p.695-708, https://doi.org/10.1002/asi.23689

Laudel, G. & Gläser, J. (2014), "Beyond breakthrough research: Epistemic properties of research and their consequences for research funding", *Research Policy 43*, p.1204-1216

Lorenz, C. (2012), "If You're So Smart, Why Are You under Surveillance? Universities, Neoliberalism, and New Public Management", *Critical Inquiry; 38(3*), p.599-629

Martinson, B.C., Anderson, M.S., Crain, A.L., De Vries, R. (2006), "Scientists' perceptions of organizational justice and self-reported misbehaviors.", *J Empir Res Hum Res Ethics*; *1(1)*: p.51–66

Martinson, B.C., Anderson, M.S., de Vries, R. "Scientists behaving badly", *Nature; 435(7043):* p.737–8

Martinson, B.C., Crain, A.L., Anderson, M.S., De Vries, R. (2009), "Institutions' Expectations for Researchers' Self-Funding, Federal Grant Holding, and Private Industry Involvement: Manifold Drivers of Self-Interest and Researcher Behavior", *Academic Medicine; 84(11)*: p.1491-1499

Martinson, B.C., Crain, L.A., De Vries. R., Anderson, M.S. (2010), "The importance of organizational justice in ensuring research integrity*", J Empir Res Hum Res Ethics;5(3):* p.67–83

Martinson, B.C., Nelson, D., Hagel-Campbell, E., Mohr, D., Charns, M.P., Bangerter, A. (2016), "Initial results from the Survey of Organizational Research Climates (SOuRCe) in the U.S. department of veterans affairs healthcare system", *PLoS One.;11(3)*: p.1–18

Martinson, B.C., Thrush, C.R., Crain, A.L. (2013), "Development and validation of the Survey of Organizational Research Climate (SORC)", *Sci Eng Ethics.;19(3)*: p.813–34

Merton, R.K. (1938), "Social structure and anomie", *American Sociological Review;*3: p.672–682

Miller, A.N., Taylor, S.G., Bedeian, A.G. (2011), "Publish or perish: academic life as management faculty live it", *Career Dev Int.;16(5)*: p.422–45

Moosa, I.A. (2018); "Publish or Perish – Perceived Benefits versus Unintended Consequences", *Edward Elgar Publishing*, https://doi.org/10.4337/9781786434937

OSTP, Federal policy on research misconduct (2000) — http://www.Ostp.Gov/html/001207_3.Html, *U.S. Office of Science and Technology Policy, Executive Office of the President*

Overman, S., Akkerman, A., Torenvlied, R. (2016), "Targets for honesty: How performance indicators shape integrity in Dutch higher education", *Public Adm.;94(4)*: p.1140–54

Porter, T. (1995), "Trust in numbers", *Princeton University Press*

Rosseel. Y. (2012), "lavaan: An R Package for Structural Equation Modeling" *Journal of Statistical Software;48(2)*, 1-36, https://www.jstatsoft.org/v48/i02/

Roy, J.R. & Mountain, M. (2006), "The Evolving Sociology of Ground-Based Optical and Infrared astronomy at the Start of the 21st Century", p.11-37, in: Heck, A. (Editor), "Organizations and Strategies in astronomy", *Volume 6, Astrophysics and Space Science Library, Springer*

Rushforth, A.D. & De Rijcke, S. (2015). "Accounting for Impact? The Journal Impact Factor and the Making of Biomedical Research in the Netherlands", *Minerva 53*, p.117-139

Siegrist, J., Li, J., Montano, D. (2014), "Psychometric properties of the Effort-Reward lmbalance Questionnaire", Department of Medical Sociology, Faculty of Medicine, Duesseldorf University, Germany

Sovacool B.K. (2008), "Exploring scientific misconduct: Isolated individuals, impure institutions, or an inevitable idiom of modern science?", *J Bioeth Inq.;5(4):* p.271–82

Stephan, P. (2012), "How economics shapes science", *Harvard University Press*

Tijdink, J.K., Schipper, K., Bouter, L.M., Pont, P.M., De Jonge, J., Smulders, Y.M. (2016), "How do scientists perceive the current publication culture? A qualitative focus group interview study among Dutch biomedical researchers", *BMJ Open; 6(2)*

Tijdink, J.K., Smulders, Y.M., Vergouwen, A.C.M., de Vet, H.C.W., Knol, D.L. (2014a), "The assessment of publication pressure in medical science; validity and reliability of a Publication Pressure Questionnaire (PPQ)", *Qual Life Res 23*: p.2055–2062 https://doi.org/10.1007/s11136-014-0643-6

Tijdink, J.K., Verbeke, R., Smulders, Y.M. (2014b), "Publication pressure and scientific misconduct in medical scientists", *J Empir Res Hum Res Ethics;9(5):* p.64–71

Tijdink, J.K., Vergouwen, A.C.M., Smulders, Y.M. (2013), "Publication pressure and burn out among Dutch medical professors: A nationwide survey", *PLoS One 3;8(9)*

Van Dalen, H.P., Henkens, K. (2012), "Intended and unintended consequences of a publish-or-perish culture: A worldwide survey", *J Am Soc Inf Sci Technol.;63(7)*: p.1282–1293

Wells, J.A., Thrush, C.R., Martinson, B.C., May, T.A., Stickler, M., Callahan, E.C., (2014), "Survey of organizational research climates in three research intensive, doctoral granting universities", *J Empir Res Hum Res Ethics;9(5)*: p.72–88

Wouters, P. (2017), "Bridging the Evaluation Gap", *Engaging Science, Technology, and Society 3*: p.108-118

Zuiderwijk, A. & Spiers, H., (2019), "Sharing and re-using open data: a case study of motivations in astrophysics", *International Journal of Information Management, Elsevier Ltd*, p.228-241, https://doi.org/10.1016/j.ijinfomgt.2019.05.024

## Supplementary Materials

**S1-Appendix: Survey Questions**

**S2-Appendix: EFAs & CFAs**

**S3-Appendix: Perceived impact of scientific misconduct on research quality**

# S1-Appendix: Survey Questions

**S1-Table1a: First item-battery for the DVs Scientific Misbehaviour and Research Quality.**

Instruction:

"Here we present 9 different forms of research behaviour. For each item, please answer the same 3 questions for the general situation in research in Astronomy as you experience it. Item 8 & 9 contain 4 questions. We are interested in your personal views and opinions. These may be based on direct experience, stories from colleagues and/or knowledge of the literature on research behaviour. Please remember that answering honestly about your personal experiences is vital for this study. Your answers are completely anonymous."

Questions for each item:

1 - How frequently does this form of research behaviour happen in Astronomy?
2 - If it occurs, how impactful is it on the validity of the findings of the study at hand?
3 - If it occurs, how impactful is it on the resulting paper's ability to convey the research appropriately?
4 - If it occurs, how impactful is it on ensuring that a diverse set of research questions are studied in Astronomy?

For each question the response scale ranged from 1=Very Low to 5=Very High.

Items:

Items were adapted to the context of astronomy based on Martinson et al. (2005, 2006, 2009, 2010) and Bouter et al. (2016). ° denotes questions added by the authors.

| Item # | Type of Misbehaviour |
|---|---|
| 1 | Inappropriate or careless peer review of papers or proposals |
| 2 | Not ensuring easy reproducibility when writing a paper |
| 3 | Spread study results over more papers than needed |
| 4 | Inadequate monitoring of research projects due to work overload |
| 5 | Cutting corners in a hurry to complete a project |
| 6 | Not sharing ancillary or meta data° |
| 7 | Not sharing the reduction algorithm used for data analysis° |
| 8 | Propose study questions solely because they are considered a 'hot' topic° |
| 9 | Not considering a study question because it isn't considered a 'hot' topic, even though it could be important for astronomy° |

**S1-Table1b: Second item-battery for the DVs Scientific Misbehaviour and Research Quality.**

Instruction:

"Here we present 9 different forms of research misbehaviour. For each item, please answer the same 3 questions for the general situation in research in Astronomy as you experience it. We are interested in your personal views and opinions. These may be based on direct experience, stories from colleagues and/or knowledge of the literature on research misbehaviour. Please remember that answering honestly about your personal experiences is vital for this study. Your answers are completely anonymous."

Questions for each item:

1 - How frequently does this form of research behaviour happen in Astronomy?
2 - If it occurs, how impactful is it on the validity of the findings of the study at hand?
3 - If it occurs, how impactful is it on the resulting paper's ability to convey the research appropriately?

For each question the response scale ranged from 1=Very Low to 5=Very High.

Items:

Items were adapted to the context of astronomy based on Martinson et al. (2005, 2006, 2009, 2010) and Bouter et al. (2016).

| **Item #** | **Type of Misbehaviour** |
|---|---|
| 10 | Giving authorship credit to someone who has not contributed substantively to a manuscript |
| 11 | Denying authorship credit to someone who has contributed substantively to a manuscript |
| 12 | Intentionally overlooking others' use of flawed data or methods |
| 13 | Data fabrication and/ or falsification |
| 14 | Compromising the rigor of a study's design or methodology in response to (publication) pressure |
| 15 | Using published ideas or phrases of others without referencing (Plagiarism) |
| 16 | Using unpublished ideas or phrases of others without their permission |
| 17 | Concealing results that contradict one's earlier findings or convictions |
| 18 | Biased interpretation of data that distorts results |

**S1-Table2: Item-battery for the IV Perceived Publication Pressure.**

Instruction:

"Please indicate to what extent you agree/ disagree with the following statements:"

Items:

Items were adapted to the context of astronomy based on Tijdink et al. (2014a). ° denotes questions added by the authors.
For each question the response scale ranged from 1=Strongly Disagree to 5=Strongly Agree and included the option "NA". * denotes reverse coded items.

| |
|---|
| Publication of scientific articles is the most important aspect of my work |
| The number of scientific publications contributes to my status |
| I experience my colleagues' assessment of me on the basis of my publications as stressful |

| I experience the publication criteria formulated by my university for my appointment or reappointment as professor as a stimulus* |
|---|
| Publication pressure puts pressure on relationships with fellow-researchers |
| In my opinion the pressure to publish scientific articles has become too high |
| The competitive scientific climate stimulates me to publish more* |
| My colleagues judge me mainly on the basis of my publications |
| In spite of the pressure to publish, I enjoy investing in other activities that I feel complement research* |
| In my experience, professors maintain their teaching skills well, despite publication pressure* |
| I cannot confide innovative research proposals to my colleagues |
| Without publication pressure, my scientific output would be of higher quality |
| My scientific publications contribute to better (future) astronomy* |
| Publication pressure results in me publishing more without it compromising the quality of my scientific work |
| I suspect that publication pressure leads some colleagues (whether intentionally or not) to fabricate data |
| The validity of astronomy literature is increased by the publication pressure in academia* |
| Publication pressure leads to serious worldwide doubts about the validity of research results |
| Publication pressure harms science |
| Without publication pressure, I would invest more time into publishing ancillary & meta data° |
| In spite of the pressure to publish, I take the time to make my reduction algorithm available and usable for other researchers°* |
| Without publication pressure I would spend more time on writing my publication in a clearly understandable and replicable way° |

| How often do you feel pressure to publish?° [This questions response scale was “Never”, “Very rarely”, “Rarely, “Regularly, “Often” and “Very often”; whereby “Never” and “Very rarely” were recoded into one category for the analysis] |
|---|

**S1-Table3: Item-battery for the IV Perceived Distributive Justice.**

Instruction:

“To what extent do you agree/ disagree with the following statements?”

Items:

Items were adapted to the context of astronomy based on Siegrist et al. (2014). ° denotes questions added by the authors. For each question the response scale ranged from 1=Strongly Disagree to 5=Strongly Agree. * denotes reverse coded items.

| **Scale** | **Item** |
|---|---|
| Effort | I have constant time pressure due to a heavy work load |
| Effort | I have many interruptions and disturbances while performing my job |
| Effort | Over the past few years, my job has become more and more demanding |
| Reward | I receive the respect I deserve from my supervisor or a respective relevant person |
| Reward | My job promotion prospects are good |
| Reward | I have experienced or I expect to experience an undesirable change in my work situation* |
| Reward | My job security is poor* |
| Reward | Considering all my efforts and achievements, I receive the respect and prestige I deserve at work |

| Reward | Considering all my efforts and achievements, my job promotion prospects are adequate |
|---|---|
| Reward | Considering all my efforts and achievements, my salary is adequate |
| Reward | My salary is fair compared to others who have similar work tasks° |

**S1-Table4a: Item-battery for the IV Perceived Organisational Justice in terms of Resource Allocation.**

Instruction:

"The following items refer to resource allocation decisions made in your primary institution that affect you. When you think of resources, think of things like salaries, office space, staff support, leaves, protected time for research, new positions, infrastructure, promotions, tenure, leadership roles and so on.
To what extent do you agree/disagree that in the past 3 years (or less if you have not been there for 3 years):"

Items:

Items were adapted to the context of astronomy based on Martinson et al. (2006). ° denotes questions added by the authors.
For each question the response scale ranged from 1=Strongly Disagree to 5=Strongly Agree. * denotes reverse coded items.

<table>
<tr><td rowspan="3"><b>Resource allocation in your institution has reflected…</b></td><td>... your effort in your work</td></tr>
<tr><td>... your contributions to the institution</td></tr>
<tr><td>... your accomplishments in your career</td></tr>
<tr><td colspan="2">I experience resource allocation in my institution as fair</td></tr>
<tr><td colspan="2">In my opinion, the allocation of resources in my institution is mostly based on improper preferential treatment°*</td></tr>
<tr><td colspan="2">In my opinion, the allocation of resources in my institution is mostly based on luck°*</td></tr>
<tr><td rowspan="3"><b>Procedures for making decisions about resource allocation in your institution have been …</b></td><td>... free of bias</td></tr>
<tr><td>... applied with consistency</td></tr>
<tr><td>... based on accurate information</td></tr>
</table>

| | |
|---|---|
| | ... Ethical |
| | ... well managed |
| | ... based on merit |
| You had an influence in these decisions | |
| You (would) have been able to appeal these decisions | |

**S1-Table4b: Item-battery for the IV Perceived Organisational Justice in terms of Peer Review.**

Instruction:

“The following items refer to the peer review of your most recent manuscript submitted for publication. When you think of the review, consider the overall quality and the review process. To what extent do you agree that:”

Items:

Items were adapted to the context of astronomy based on Martinson et al. (2006). ° denotes questions added by the authors.
For each question the response scale ranged from 1=Strongly Disagree to 5=Strongly Agree. * denotes reverse coded items.

| | |
|---|---|
| **The reviews were appropriate relative to …** | ... the effort you put into the manuscript |
| | ... the quality of the manuscript |
| I experience the reviews as fair | |
| In my opinion, the acceptance or refusal of a manuscript is mostly based on improper preferential treatment°* | |
| In my opinion, the acceptance or refusal of a manuscript is mostly based on luck°* | |
| | ... free of bias |
| | ... applied with consistency |
| | ... based on accurate information |
| **The review process was …** | ... Ethical |

| | ... well managed |
|---|---|
| | ... based on merit |
| You (would) have been able to appeal the review decision | |
| The review process was typical of reviews you have received in the past 3 years | |

**S1-Table4c: Item-battery for the IV Perceived Organisational Justice in terms of Grant Application.**

Instruction:

"The following items refer to the review of your most recent extramural grant application. When you think of the review, consider the overall quality of the review and the review process. To what extent do you agree that:"

Items:

Items were adapted to the context of astronomy based on Martinson et al. (2006). ° denotes questions added by the authors.
For each question the response scale ranged from 1=Strongly Disagree to 5=Strongly Agree. * denotes reverse coded items.

| **The reviews were appropriate relative to …** | ... the effort you put into the application |
|---|---|
| | ... the quality of your proposal |
| | ... your research track record |
| I experience the reviews as fair | |
| In my opinion, decisions on funding are mostly based on improper preferential treatment°* | |
| In my opinion, decisions on funding are mostly based on luck°* | |
| **The review process was …** | ... free of bias |
| | ... applied with consistency |
| | ... based on accurate information |
| | ... Ethical |
| | ... well managed |
| | ... based on merit |
| You (would) have been able to appeal the review decision | |
| The review process was typical of reviews you have received in the past 3 years | |

**S1-Table4d: Item-battery for the IV Perceived Organisational Justice in terms of Telescope Time Application.**

Instruction:

"The following items refer to the review of your latest application for telescope time. When you think of the review, consider the overall quality of the review and the review process. To what extent do you agree that:"

Items:

Items were adapted to the context of astronomy based on Martinson et al. (2006). ° denotes questions added by the authors.
For each question the response scale ranged from 1=Strongly Disagree to 5=Strongly Agree. * denotes reverse coded items.

| | |
|---|---|
| **The reviews were appropriate relative to …** | ... the effort you put into the application |
| | ... the quality of your proposal |
| | ... your research track record |
| I experience the reviews as fair | |
| In my opinion, the allocation of telescope is mostly based on improper preferential treatment°* | |
| In my opinion, the allocation of telescope is mostly based on luck°* | |
| **The review process was …** | ... free of bias |
| | ... applied with consistency |
| | ... based on accurate information |
| | ... Ethical |
| | ... well managed |
| | ... based on merit |
| You (would) have been able to appeal the review decision | |
| The review process was typical of reviews you have received in the past 3 years | |

**S1-Table5: Item-battery for the IV Perceived Overcommitment.**

Instruction:

"To what extent do you agree/ disagree with the following statements?"

Items:

Items were adapted to the context of astronomy based on Siegrist et al. (2014).
For each question the response scale ranged from 1=Strongly Disagree to 5=Strongly Agree. * denotes reverse coded items.

| |
|---|
| I get easily overwhelmed by time pressures at work |
| As soon as I get up in the morning I start thinking about work |
| After I finish my work day, I can easily relax and 'switch off' work* |
| People close to me say I sacrifice too much for my job |
| Work rarely lets me go, it is still on my mind when I go to bed |
| If I postpone something at work that I was supposed to do today I'll have trouble sleeping at night |

## S2-Appendix: EFAs & CFAs

### S2-Table1a: PPQ Cronbach Alphas before removal of items

This table displays the SPSS output of the Cronbach alphas of the PPQ consisting of 22 items (see *S1-Appendix: S1-Table2*). * denotes items that we subsequently removed.

**Reliability Statistics**

| Cronbachs Alpha | N of Items |
|---|---|
| .768 | 22 |

**Item-Total-Statistics**

| PPQ Items | Scale Mean if Item Deleted | Scale Variance if Item Deleted | Corrected Item-Total Correlation | Cronbach's Alpha if Item Deleted |
|---|---|---|---|---|
| Publication of scientific articles is the most important aspect of my work* | 64.87 | 107.316 | -.012 | .780 |
| The number of scientific publications contributes to my status* | 64.21 | 104.431 | .162 | .768 |
| I experience my colleagues' assessment of me on the basis of my publications as stressful | 64.99 | 93.519 | .541 | .744 |
| I experience the publication criteria formulated by my university for my appointment or reappointment as professor as a stimulus* | 64.89 | 107.312 | -.015 | .781 |
| Publication pressure puts pressure on relationships with fellow-researchers | 64.82 | 93.923 | .545 | .744 |
| In my opinion the pressure to publish scientific articles has become too high | 64.19 | 94.771 | .573 | .744 |
| The competitive scientific climate stimulates me to publish more* | 65.22 | 105.804 | .042 | .778 |
| My colleagues judge me mainly on the basis of my publications | 65.02 | 99.676 | .317 | .760 |
| In spite of the pressure to publish, I enjoy investing in other activities that I feel complement research* | 66.21 | 107.590 | -.012 | .778 |
| In my experience, professors maintain their teaching skills well, despite publication pressure* | 65.08 | 101.890 | .230 | .765 |
| I cannot confide innovative research proposals to my colleagues | 65.71 | 97.539 | .401 | .754 |
| Without publication pressure, my scientific output would be of higher quality | 65.10 | 92.207 | .605 | .739 |
| My scientific publications contribute to better (future) astronomy* | 66.14 | 105.698 | .124 | .769 |
| Publication pressure results in me publishing more without it compromising the quality of my scientific work* | 65.47 | 110.134 | -.131 | .786 |
| I suspect that publication pressure leads some colleagues (whether intentionally or not) to fabricate data | 65.33 | 96.520 | .424 | .753 |
| The validity of astronomy literature is increased by the publication pressure in academia* | 64.55 | 103.824 | .162 | .769 |
| Publication pressure leads to serious worldwide doubts about the validity of research results | 64.87 | 94.650 | .534 | .745 |
| Publication pressure harms science | 64.51 | 93.764 | .580 | .742 |
| Without publication pressure, I would invest more time into publishing ancillary & meta data | 65.05 | 94.540 | .523 | .746 |
| In spite of the pressure to publish, I take the time to make my reduction algorithm available and usable for other researchers* | 65.57 | 103.527 | .149 | .771 |
| Without publication pressure I would spend more time on writing my publication in a clearly understandable and replicable way | 65.13 | 92.224 | .577 | .741 |
| How often do you feel pressure to publish? | 65.03 | 94.954 | .555 | .745 |

**S2-Table1b: PPQ Cronbach Alphas after removal of items**

This table displays the SPSS output of the Cronbach alphas of the cleaned PPQ after removal of items, which resulted in 12 remaining items.

**Reliability Statistics**

| Cronbachs Alpha | Cronbachs Alpha for standardised Items | N of Items |
|---|---|---|
| .871 | .871 | 12 |

**Item-Total-Statistics**

| Cleaned PPQ Items | Scale Mean if Item Deleted | Scale Variance if Item Deleted | Corrected Item-Total Correlation | Squared Multiple Correlation | Cronbach's Alpha if Item Deleted |
|---|---|---|---|---|---|
| I experience my colleagues' assessment of me on the basis of my publications as stressful | 34.51 | 73.685 | .589 | .424 | .859 |
| Publication pressure puts pressure on relationships with fellow-researchers | 34.36 | 73.539 | .620 | .419 | .857 |
| In my opinion the pressure to publish scientific articles has become too high | 33.66 | 75.354 | .610 | .465 | .858 |
| My colleagues judge me mainly on the basis of my publications | 34.57 | 79.577 | .356 | .225 | .873 |
| I cannot confide innovative research proposals to my colleagues | 35.27 | 78.166 | .427 | .237 | .869 |
| Without publication pressure, my scientific output would be of higher quality | 34.63 | 72.761 | .655 | .530 | .854 |
| I suspect that publication pressure leads some colleagues (whether intentionally or not) to fabricate data | 34.82 | 77.464 | .426 | .237 | .869 |
| Publication pressure leads to serious worldwide doubts about the validity of research results | 34.37 | 75.231 | .566 | .413 | .860 |
| Publication pressure harms science | 33.97 | 75.126 | .595 | .497 | .859 |
| Without publication pressure, I would invest more time into publishing ancillary & meta data | 34.57 | 74.144 | .600 | .432 | .858 |
| Without publication pressure I would spend more time on writing my publication in a clearly understandable and replicable way | 34.71 | 72.160 | .656 | .547 | .854 |
| How often do you feel pressure to publish? | 34.59 | 74.540 | .588 | .392 | .859 |

**S2-Table1c: CATPCA for cleaned PPQ**

**Pattern Matrix**

| Cleaned PPQ Items | **Dimension** | | |
|---|---|---|---|
| | F1 (Extent & Consequences on one's own conduct of research) | F2 (Impact on Relationship with colleagues) | F3 (Suspected consequences on science) |
| Without publication pressure I would spend more time on writing my publication in a clearly understandable and replicable way | .811 | .008 | -.022 |
| Without publication pressure, my scientific output would be of higher quality | .809 | -.058 | .057 |
| Without publication pressure, I would invest more time into publishing ancillary & meta data | .779 | .001 | -.073 |
| Publication pressure harms science | .639 | -.138 | .328 |
| In my opinion the pressure to publish scientific articles has become too high | .585 | .115 | .167 |
| How often do you feel pressure to publish? | .522 | .521 | -.227 |
| My colleagues judge me mainly on the basis of my publications | -.169 | .866 | -.060 |
| I experience my colleagues' assessment of me on the basis of my publications as stressful | .298 | .662 | -.099 |
| I cannot confide innovative research proposals to my colleagues | -.300 | .595 | .506 |
| [Publication pressure puts pressure on relationships with fellow-researchers | .167 | .552 | .240 |
| I suspect that publication pressure leads some colleagues (whether intentionally or not) to fabricate data | .005 | -.004 | .799 |
| Publication pressure leads to serious worldwide doubts about the validity of research results | .368 | -.103 | .632 |

**S2-Table2a: CATPCA for ERI**

**Pattern Matrix**

| ERI Items | **Dimension**<br>F5 (Overcommitment) | F1 (Reward: Job situation) | F3 (Effort) | F2 (Reward: Salary) | F4 (Reward: Receiving praise/ respect) |
|---|---|---|---|---|---|
| Work rarely lets me go, it is still on my mind when I go to bed | .850 | .031 | -.018 | .044 | -.095 |
| As soon as I get up in the morning I start thinking about work | .838 | .090 | -.104 | .013 | .018 |
| After I finish my work day, I can easily relax and 'switch off' work | .736 | .019 | .004 | .011 | -.147 |
| If I postpone something at work that I was supposed to do today I'll have trouble sleeping at night | .651 | -.112 | .016 | .074 | -.020 |
| People close to me say I sacrifice too much for my job | .494 | .010 | .222 | -.203 | .029 |
| My job security is poor | -.094 | .906 | .107 | -.025 | -.279 |
| My job promotion prospects are good | .156 | .689 | .041 | .077 | .171 |
| I have experienced or I expect to experience an undesirable change in my work situation | -.021 | .658 | -.181 | -.231 | .151 |
| Considering all my efforts and achievements, my job promotion prospects are adequate | .071 | .631 | .027 | .252 | .167 |
| I have many interruptions and disturbances while performing my job | -.219 | .145 | .829 | .106 | -.258 |
| I have constant time pressure due to a heavy work load | .144 | .034 | .740 | -.055 | .097 |
| Over the past few years, my job has become more and more demanding | .060 | -.002 | .693 | -.170 | .193 |
| I get easily overwhelmed by time pressures at work | .113 | -.256 | .563 | .108 | .074 |
| Considering all my efforts and achievements, my salary is adequate | .022 | .014 | -.005 | .917 | -.019 |
| My salary is fair compared to others who have similar work tasks | .008 | -.047 | -.016 | .888 | .069 |
| I receive the respect I deserve from my supervisor or a respective relevant person | -.082 | -.077 | .013 | -.017 | .907 |
| Considering all my efforts and achievements, I receive the respect and prestige I deserve at work | -.107 | .155 | .020 | .113 | .721 |

**S2-Table2b: Cronbach Alphas for Effort Construct**

**Reliability Statistics**

| Cronbachs Alpha | N of Items |
|---|---|
| .691 | 4 |

**Item-Total-Statistics**

| ERI F3 Items (Effort) | Scale Mean if Item Deleted | Scale Variance if Item Deleted | Corrected Item-Total Correlation | Cronbach's Alpha if Item Deleted |
|---|---|---|---|---|
| I have constant time pressure due to a heavy work load | 10.51 | 6.109 | .616 | .532 |
| I have many interruptions and disturbances while performing my job | 10.66 | 7.295 | .364 | .694 |
| Over the past few years, my job has become more and more demanding | 10.46 | 6.900 | .490 | .618 |
| I get easily overwhelmed by time pressures at work | 11.05 | 6.691 | .443 | .648 |

**S2-Table2c: Cronbach Alphas for Overcommitment Construct**

**Reliability Statistics**

| Cronbachs Alpha | N of Items |
|---|---|
| .780 | 5 |

**Item-Total-Statistics**

| ERI F5 Items (Overcommitment) | Scale Mean if Item Deleted | Scale Variance if Item Deleted | Corrected Item-Total Correlation | Cronbach's Alpha if Item Deleted |
|---|---|---|---|---|
| As soon as I get up in the morning I start thinking about work | 13.36 | 13.876 | .536 | .745 |
| After I finish my work day, I can easily relax and 'switch off' work | 13.62 | 12.880 | .577 | .731 |
| People close to me say I sacrifice too much for my job | 13.95 | 13.299 | .493 | .761 |
| Work rarely lets me go, it is still on my mind when I go to bed | 13.52 | 12.527 | .683 | .697 |
| If I postpone something at work that I was supposed to do today I'll have trouble sleeping at night | 14.21 | 13.216 | .499 | .759 |

**S2-Table2d: Cronbach Alphas for Reward Construct**

**Reliability Statistics**

| Cronbachs Alpha | N of Items |
| --- | --- |
| .805 | 8 |

**Item-Total-Statistics**

| ERI F1+F2+F4 Items (Reward) | Scale Mean if Item Deleted | Scale Variance if Item Deleted | Corrected Item-Total Correlation | Cronbach's Alpha if Item Deleted |
| --- | --- | --- | --- | --- |
| I receive the respect I deserve from my supervisor or a respective relevant person | 21.90 | 37.123 | .451 | .793 |
| My job promotion prospects are good | 22.96 | 35.394 | .564 | .777 |
| I have experienced or I expect to experience an undesirable change in my work situation | 22.39 | 36.035 | .431 | .797 |
| My job security is poor | 22.31 | 34.700 | .413 | .805 |
| Considering all my efforts and achievements, I receive the respect and prestige I deserve at work | 22.33 | 35.336 | .627 | .770 |
| Considering all my efforts and achievements, my job promotion prospects are adequate | 22.76 | 33.467 | .685 | .759 |
| Considering all my efforts and achievements, my salary is adequate | 22.53 | 34.976 | .520 | .783 |
| My salary is fair compared to others who have similar work tasks | 22.43 | 35.253 | .520 | .783 |

**S2-Table3: Results of the comparative factor analyses (CFAs) for all independent variables**

| **Independent Variable** | **Comparative Fit Index (CFI)** | **Tucker-Lewis Index (TLI)** | **Root Mean Square Error of Approximation (RMSEA)** | **Difference to model without accounting for covariances between indicators** |
|---|---|---|---|---|
| Publication Pressure (PPQ) | 0.958 | 0.939 | 0.063 90%CI(.057, .069). | $\chi 2(5)= 374.19$; $p<.001$ |
| Effort Reward Imbalance (ERI) | 0.967 | 0.953 | 0.044 90%CI(.039, .048) | $\chi 2(18)= 475.02$; $p<.001$ |
| Organisational Justice: Resource Allocation | 0.977 | 0.970 | 0.055 90%CI(.050, .059) | $\chi 2(8)= 3655.6$; $p<.001$ |
| Organisational Justice: Peer Review | 0.969 | 0.960 | 0.064 90%CI(.060, .069) | $\chi 2(5)= 1708.7$; $p<.001$ |
| Organisational Justice: Grant Application | 0.979 | 0.971 | 0.055 90%CI(.049, .061) | $\chi 2(10)= 1220.6$; $p<.001$ |
| Organisational Justice: Telescope Time Application | 0.963 | 0.952 | 0.068 90%CI(0.062, 0.074) | $\chi 2(7)= 982.7$; $p<.001$ |

# S3-Appendix: Perceived impact of scientific misconduct on research quality

**S3-Table1: Perceived frequency of scientific misconduct and impact thereof on research quality**

This table displays the means of perceived frequency and impact on the respective quality criterion of each misbehaviour item (scale range: 1-5). The rank numbers are calculated in descending order. QC1 (Quality Criterion 1): "validity of the findings at hand"; QC2 (Quality Criterion 2): "the resulting paper's ability to convey the research appropriately"; QC3 (Quality Criterion 3): "impact on research diversity"

| Item # | Type of Misbehaviour | Frequency | Frequency Rank | QC1 | QC1 Rank | QC2 | QC2 Rank | QC3 | QC3 Rank |
|---|---|---|---|---|---|---|---|---|---|
| 1 | Inappropriate or careless peer review of papers or proposals | 2.76 | 12 | 3.31 | 9 | 3.35 | 9 | | |
| 2 | Not ensuring easy reproducibility when writing a paper | 3.19 | 9 | 3.51 | 7 | 3.46 | 7 | | |
| 3 | Spread study results over more papers than needed | 3.30 | 6 | 2.61 | 16 | 2.99 | 15 | | |
| 4 | Inadequate monitoring of research projects due to work overload | 3.42 | 3 | 3.40 | 8 | 3.37 | 8 | | |
| 5 | Cutting corners in a hurry to complete a project | 3.23 | 8 | 3.66 | 6 | 3.52 | 6 | | |
| 6 | Not sharing ancillary or meta data | 3.29 | 7 | 3.29 | 10 | 3.31 | 10 | | |
| 7 | Not sharing the reduction algorithm used for data analysis | 3.40 | 5 | 3.22 | 11 | 3.21 | 11 | | |
| 8 | Propose study questions solely because they are considered a 'hot' topic | 3.88 | 1 | 2.91 | 13 | 2.87 | 16 | 3.70 | 2 |
| 9 | Not considering a study question because it isn't considered a 'hot' topic, even though it could be important for astronomy | 3.40 | 4 | 3.05 | 12 | 3.08 | 13 | 3.79 | 1 |
| 10 | Giving authorship credit to someone who has not contributed substantively to a manuscript | 3.73 | 2 | 1.97 | 18 | 2.06 | 18 | | |
| 11 | Denying authorship credit to someone who has contributed substantively to a manuscript | 2.05 | 17 | 2.53 | 17 | 2.67 | 17 | | |
| 12 | Intentionally overlooking others' use of flawed data or methods | 2.48 | 16 | 3.74 | 5 | 3.61 | 5 | | |
| 13 | Data fabrication and/ or falsification | 1.95 | 18 | 4.17 | 1 | 4.01 | 1 | | |
| 14 | Compromising the rigor of a study's design or methodology in response to (publication) pressure | 2.93 | 11 | 3.80 | 4 | 3.64 | 4 | | |
| 15 | Using published ideas or phrases of others without referencing (Plagiarism) | 2.61 | 13 | 2.82 | 14 | 3.18 | 12 | | |
| 16 | Using unpublished ideas or phrases of others without their permission | 2.49 | 15 | 2.80 | 15 | 3.07 | 14 | | |
| 17 | Concealing results that contradict one's earlier findings or convictions | 2.59 | 14 | 3.93 | 3 | 3.84 | 3 | | |
| 18 | Biased interpretation of data that distorts results | 3.08 | 10 | 4.04 | 2 | 3.95 | 2 | | |

**S3-Table2: Perceived impact of scientific misbehavior on research quality**

This table displays the perceived impact for each of 18 types of misbehaviour as the product score of the means of perceived frequency of each misbehaviour and impact on the three aspects of research quality (numbers taken from *S3-Table1*). The rank number is calculated in descending order. Given that the answer scales range from 1 to 5; the results of perceived impact can range between 1 and 25.

| Type of misbehaviour & Type of quality criterion | Perceived impact | Rank Number | Type of misbehaviour & Type of quality criterion | Perceived impact | Rank Number |
|---|---|---|---|---|---|
| "Propose study questions solely because they are considered a 'hot' topic" & QC3 | 14.33 | 1 | "Not considering a study question because it isn't considered a 'hot' topic, even though it could be important for astronomy" & QC1 | 10.36 | 20 |
| "Not considering a study question because it isn't considered a 'hot' topic, even though it could be important for astronomy" & QC3 | 12.89 | 2 | "Concealing results that contradict one's earlier findings or convictions" & QC1 | 10.18 | 21 |
| "Biased interpretation of data that distorts results" & QC1 | 12.47 | 3 | "Concealing results that contradict one's earlier findings or convictions" & QC2 | 9.95 | 22 |
| "Biased interpretation of data that distorts results" & QC2 | 12.17 | 4 | "Spread study results over more papers than needed" & QC2 | 9.87 | 23 |
| "Cutting corners in a hurry to complete a project" & QC1 | 11.80 | 5 | "Intentionally overlooking others' use of flawed data or methods" & QC1 | 9.27 | 24 |
| "Inadequate monitoring of research projects due to work overload" & QC1 | 11.61 | 6 | "Inappropriate or careless peer review of papers or proposals" & QC2 | 9.24 | 25 |
| "Inadequate monitoring of research projects due to work overload" & QC2 | 11.51 | 7 | "Inappropriate or careless peer review of papers or proposals" & QC1 | 9.14 | 26 |
| "Cutting corners in a hurry to complete a project" & QC2 | 11.37 | 8 | "Intentionally overlooking others' use of flawed data or methods" & QC2 | 8.97 | 27 |
| "Propose study questions solely because they are considered a 'hot' topic" & QC1 | 11.29 | 9 | "Spread study results over more papers than needed" & QC1 | 8.61 | 28 |
| "Not ensuring easy reproducibility when writing a paper" & QC1 | 11.2 | 10 | "Using published ideas or phrases of others without referencing (Plagiarism)" & QC2 | 8.29 | 29 |
| "Compromising the rigor of a study's design or methodology in response to (publication) pressure" & QC1 | 11.14 | 11 | "Data fabrication and/ or falsification" & QC1 | 8.15 | 30 |
| “Propose study questions solely because they are considered a 'hot' topic” & QC2 | 11.12 | 12 | “Data fabrication and/ or falsification” & QC2 | 7.84 | 31 |
| “Not ensuring easy reproducibility when writing a paper” & QC2 | 11.03 | 13 | “Giving authorship credit to someone who has not contributed substantively to a manuscript” & QC2 | 7.67 | 32 |
| “Not sharing the reduction algorithm used for data analysis” & QC1 | 10.93 | 14 | “Using unpublished ideas or phrases of others without their permission” & QC2 | 7.65 | 33 |
| “Not sharing the reduction algorithm used for data analysis” & QC2 | 10.92 | 15 | “Using published ideas or phrases of others without referencing (Plagiarism)” & QC1 | 7.37 | 34 |
| “Not sharing ancillary or meta data” & QC2 | 10.91 | 16 | “Giving authorship credit to someone who has not contributed substantively to a manuscript” & QC1 | 7.34 | 35 |
| “Not sharing ancillary or meta data” & QC1 | 10.82 | 17 | “Using unpublished ideas or phrases of others without their permission” & QC1 | 6.96 | 36 |
| “Compromising the rigor of a study’s design or methodology in response to (publication) pressure” & QC2 | 10.66 | 18 | “Denying authorship credit to someone who has contributed substantively to a manuscript” & QC2 | 5.48 | 37 |
| “Not considering a study question because it isn't considered a 'hot' topic, even though it could be important for astronomy” & QC2 | 10.47 | 19 | “Denying authorship credit to someone who has contributed substantively to a manuscript” & QC1 | 5.19 | 38 |